\documentclass[fleqn,usenatbib]{mnras}
\usepackage{newtxtext,newtxmath}
\usepackage[T1]{fontenc}
\usepackage{ae,aecompl}

\usepackage{xcolor}
\usepackage{ragged2e}

\usepackage{subfig,graphicx,caption}
\usepackage{amsmath}	    

\graphicspath{{./plots/}}

\title[Simulating Cosmological Microlensing and Caustic Crossing Events]
{An Efficient Method for Simulating Light Curves of Cosmological Microlensing and Caustic Crossing Events}

\author[A. K. Meena, O. Arad, and A. Zitrin]{
Ashish Kumar Meena$^{1}$\thanks{E-mail: ashishmeena766@gmail.com},
Ofir Arad$^{1}$, and
Adi Zitrin$^{1}$
\\
\\
$^{1}$Physics Department, Ben-Gurion University of the Negev, P.O. Box 653,
Be'er-Sheva 8410501, Israel}

\pubyear{2021}

\begin{document}
\label{firstpage}
\pagerange{\pageref{firstpage}--\pageref{lastpage}}
\maketitle

\begin{abstract}
A new window to observing individual stars and other small sources at cosmological distances was opened recently, with the detection of several caustic-crossing events in galaxy cluster fields. Many more such events are expected soon from dedicated campaigns with the \emph{Hubble Space Telescope} and from the \emph{James Webb Space Telescope}. These events can teach us not only about the lensed sources themselves, such as individual high-redshift stars, star clusters, or accretion disks, but through their light-curves they also hold information about the point-mass function of the lens and thus, potentially, the composition of dark matter. We present here a simple method for simulating light curves of such events, i.e., the change in apparent magnitude of the source as it sweeps over the net of caustics generated by microlenses embedded around the critical region of the lens. The method is recursive and so any reasonably sized small source can be accommodated, down to sub-solar scales, in principle. We compare the method, which we dub \emph{Adaptive Boundary Method}, with other common methods such as simple inverse ray shooting, and demonstrate that it is significantly more efficient and accurate in the small-source and high-magnification regime of interest. A \textsc{python} version of the code is made publicly available in an open-source fashion for simulating future events.\\
\end{abstract}


\section{Introduction}
\label{sec:Introduction}

Light rays from background objects are deflected when they pass near massive bodies; a phenomenon known as gravitational lensing 
\citep[e.g.,][]{1992grle.book.....S, 1996astro.ph..6001N}. 

Lensing by galaxy clusters has played a key role in obtaining various fundamental results in astronomy. These include, for example, confirming the high mass-to-light ratios in galaxy 
clusters, indicative of a ``missing mass" component  \citep[as originally suggested by][see e.g., \citealt{Bergmann1990ApJ...350...23B,Hoekstra2002MNRAS.333..911H,Bahcall2014MNRAS.439.2505B}]{Zwicky1937MissingMass}; constraining the self-interaction cross section of 
dark matter from bullet clusters \citep[e.g.,][]{Markevitch2004, Bradac2006Bullet,Clowe2006Bullet}; detecting record breaking high-redshift galaxies \cite[e.g.,][]{Franx1997highz, 
Kneib2004z7, Bradley2008highz, Zheng2012Nature, Coe2012highz, Zitrin2014highz}; and correspondingly, 
constraining the high-redshift galaxy luminosity function and its implications for reionization \citep[e.g.,][]{
Oesch2014LumZ910, 2015ApJ...800...18A, Mcleod2016, Livermore2016LF, 2019MNRAS.486.3805B}.

Lensing by individual galaxies has also played key roles in our understanding of the cosmos. Given the shorter time-delays between multiple images in galaxy lenses compared to 
cluster lenses (although see \citealt{Kelly2016reappearance,Grillo2018H0m1149}), strongly lensed quasars have provided competitive constraints on the 
Hubble constant \citep[e.g.,][]{Suyu2010MeasuredH, 2020MNRAS.498.1420W, 2020A&A...639A.101M, 
2020MNRAS.494.6072S}, as well as constraints on the composition of dark matter from flux 
anomalies and microlensing fluctuations \citep[e.g.,][]{Mediavilla2017DMsubhaloes, 
2020A&A...633A.107H}, and on the stellar initial mass function \citep[IMF; e.g.,][]{2016MNRAS.459.3677L,
2017ApJ...845..157N}.

A few years ago, a new window to observing individual stars (and potentially other small sources) at cosmological distances was opened with the serendipitous detection of a lensed star at a redshift of $z=1.49$ behind the galaxy cluster MACS1149 \citep{Kelly2017CC}. The observed star sits in a spiral galaxy multiply imaged by the massive galaxy cluster. A small part of the galaxy sits atop the caustic so that it is highly magnified. As stars sweep across the caustics due mainly to transverse relative motion between the lens and source, they get extremely magnified and thus observed as transient phenomena \citep[e.g.,][]{Miralda-Escude1991,Diego2017CC,Venumadhav2017}. These events are sometimes called \emph{caustic transients}, or \emph{caustic crossing events} (CCEs). Several other such events have been detected since \textcolor{red}{\citep[][]{Rodney2018Trans0416,Chen2019MACS0416CCE, Kaurov2019MACS0416CCE, 2022Natur.603..815W}}, and many more are expected soon with the \emph{Hubble Space Telescope} and the \emph{James Webb Space Telescope}  \citep[e.g.,][]{Windhorst2018,Oguri2017CausticCrossing}.

In gravitational lensing theory, the critical curves and caustics mark the regions of very-high (and in the geometric optics regime, in principle, infinite) magnification in the lens and source plane, respectively \citep[e.g.,][]{1992grle.book.....S}. 
Images forming near or across the critical curves correspond to a source lying near or on the caustics in the source plane. Such a source will thus be highly magnified.
The maximum magnification value that a source can attain is governed (or limited) by: 
(i) the size of the source: the smaller the source, the higher the magnification
\citep[e.g.,][]{1987A&A...171...49S}, 
(ii) diffraction effects~\citep[e.g.,][]{1983ApJ...271..551O}, and 
(iii) the presence of microlenses in the lens~\citep[e.g.,][]{1990LNP...360..186W}.
For a smooth lens, a relatively small galaxy-scale source lying on the (macro-) caustic can have maximum magnification factors of order ${\sim}10^2$. A stellar source, in principle, can reach magnification factors up to $10^6-10^7$ (in particular, in $\sim$optical wavelengths). In reality, the macro-lens is filled with micro-lenses such as stars, black holes or other compact objects, which break the smooth macro-caustic into a net of micro-caustics, and lower the peak magnifications. As a result, the typical light curve of a background star near a caustic would exhibit microlensing fluctuations corresponding to the point-mass composition of the lens. Our goal here is to supply an efficient method for simulating these. Such light curves are important to have (and compare with observations) because they hold information on the source size and point mass population in the lens \citep{Diego2017CC,Oguri2017CausticCrossing,Venumadhav2017}. 

The phenomenon referred to herein as CCE (and nearby microlensing fluctuations) is similar in principle to the well-known microlensing fluctuations in the light of distant lensed quasars. In both cases, the source crosses the caustic structure generated by the main (macro) lens and the microlenses within it. Methods to simulate quasar microlensing light-curves have been long developed \citep[e.g.,][]{Wambsganss1990QuasarMicorlensing,Mediavilla2006IPM}. Various semi-analytic approximation has been also put forward to estimate the magnification distribution in presence of microlensing \citep[e.g.,][]{1991A&A...251..393M, 1991ApJ...374...83R, 1997ApJ...489..508K}. However, these approximations are not suitable for microlensing in very high magnifications \citep[e.g.,][]{2020PDU....2900567F}. Hence, the much smaller source size and higher magnification involved in CCEs call for a somewhat different framework, that may be more efficient in simulating the relevant light curves.

A typical size of a quasar is ${\sim}10^{16}$ -- ${\sim}10^{17}$ cm (for, say, the broad line region, see e.g., \citealt{Chelouche2012ApJ...750L..43C, Chartas2016AN....337..356C,Fian2018ApJ...859...50F}), whereas a stellar source would have typical sizes in the range ${\sim} 10^{10}$ -- ${\sim} 10^{13}$ cm. Such a difference in source size leads to orders of magnitude difference in the maximum magnification that these sources can have. Also, due to the larger size of the quasar, it covers numerous micro-caustics in the source plane leading to smoother features in the corresponding light curve. In contrast, a stellar source would show sharp features in its light curve, as it only covers a few micro-caustics at a time~\citep[e.g.,][]{Venumadhav2017, Diego2017CC}. 
In general, the macro-magnification in quasar lensing and stellar CCEs is also, often, significantly different.  The macro-magnification in quasar lensing is usually about ${\sim}10$~\citep[e.g.,][]{2019MNRAS.483.4242L, 2020MNRAS.494.3491L} whereas the macro-magnification in  CCEs is ${\gg}100$~\citep[e.g.,][]{Kelly2017CC}, as the source typically lies nearer to the macro-caustic. Such a high macro-magnification in CCEs is required in order to observe the distant stellar objects as they are inherently very faint.

In addition, due to the larger source size and the decades-long monitoring time typically needed, a large field of view (FOV) is usually required to study the corresponding light curves in quasar microlensing, for which tree methods, for example, can be advantageous
\citep[e.g.,][]{Kayser1986, Wambsganss1990QuasarMicorlensing, 2007MNRAS.381.1591B}.  For such cases, in order to simulate many light curves it makes sense to model some large image plane region and then simply ``draw" source paths across the resulting caustic patterns, convolving them with the source shape. However, for small sources such as stars, the number of caustics covered by the source in each time step is much smaller, but the effective pixel scale must be small enough to model sufficiently accurately the magnification across the source. Therefore, in such cases it may be beneficial to model smaller source plane areas, concentrating on a particular source path per run rather than constructing the full source-plane 2D map in very high resolution.

We present here a simple and efficient method for calculating light curves for CCEs and nearby microlensing fluctuations, of small sources. In recent years several works have simulated CCE light curves, and investigated what constraints could be placed on dark matter using CCEs \citep[e.g.,][]{Diego2017CC,Oguri2017CausticCrossing,Venumadhav2017,Dai2019axionCCE}. We present here our new, adaptive boundary method (ABM), which is designed specifically to handle small sources. We make our pipeline publicly available\footnote{\url{https://github.com/akmeena766/gl_cce}} in the form of an open source, free to use by the community to simulate such events.

This paper is organised as follows. In Section~\ref{sec:abm}, we present the details of our ABM code, whereas in Section~\ref{sec:comparison}, we use it for microlensing and CCE light curve simulations. We start with a source near -- but sufficiently far from -- the caustic, approximating the macrolens contribution as constant convergence and shear. We simulate various microlensing light curves using both the ABM and inverse ray-shooting (IRS) methods, and compare their performance. In Section~\ref{sec:cc_lc}, as an extreme-magnification test case for the method, we then place the source nearer to (-- essentially atop) the caustic of an input macrolens, and use ABM to simulate actual CCE light curves, both in the absence and in the presence of microlenses. To put our simulations in observational context, in Section ~\ref{sec:mu_stellar} we then estimate the minimum magnification that is needed for different stars at cosmological distances to be observed in different HST and JWST bands, for various depths. The relevant code for this calculation is also made publicly available together with the main pipeline. The work is summarised in Section~\ref{sec:conclusions}.

Throughout this work, the lens redshift ($z_{\rm L}$) and source redshift ($z_{\rm S}$)  are fixed to 0.5 and 1.5, respectively (although these can be of course set to any relevant value when running the code).
The cosmological parameters used in this work to calculate the various quantities are: 
$H_0 {=} 70 \: {\rm km} \: {\rm s}^{-1} \: {\rm Mpc}^{-1}$, $\Omega_\Lambda {=} 0.7$, and 
$\Omega_m {=} 0.3$.\\

\section{Adaptive Boundary Method (ABM)}
\label{sec:abm}

In this section we describe a simple and efficient method for calculating CCEs and nearby microlensing light curves. The method, ABM, is based on the adaptive refinement of lens and source plane regions in order to minimise the number of unnecessary operations, speeding up the calculation. In essence, the idea is as follows. The relevant image-plane area is pixelated and the center of each pixel is mapped back to the source plane using the lens equation. Then, only the pixels that mapped back to within some distance from the source position are further refined, while with each iteration the allowed distance in the source-plane is lowered as well. This process goes on until we reach the required source size, essentially guaranteeing sufficient sampling resolution. As we only focus on the relevant pixels in the image plane, we show that ABM turns out to be more efficient and more accurate than basic or benchmark methods such as IRS. In contrast, the ABM is not designed to calculate the entire caustic map in high-resolution. Instead, it is designed to produce individual light curves per run. This helps us to efficiently reach very high resolutions compared to traditional IRS \citep[e.g.,][]{Kayser1986}.

Indeed, other methods, e.g., for quasar microlensing, that are more efficient than IRS exist as well, perhaps most notably the Inverse Polygon Method \citep[IPM;][]{Mediavilla2006IPM,Mediavilla2011ApJ...741...42M} or newer methods that rely on it \citep[e.g.,][]{Shalyapin2021A&A...653A.121S}. These methods have proven excellent for the relatively-low magnification regime relevant for quasar microlensing. However, it is not clear if they would be optimal for the extreme-magnification regime, because there the magnification (and parity) can significantly change within and between adjacent cells, leading again to smaller and smaller cells and possibly complicating the calculation. Another efficient method proposed in the past is the contour method, discussed in \citep{1993MNRAS.261..647L, WyitheWebster1999Contours}, which keeps track of each lensed image formed in the image-plane and treats the light curves as a line.
Recently, \citet{2022arXiv220505141D} introduced a new algorithm for fast simulation of light curves in high magnification regions. The underlying method relies on the smoothness of the deflection field and its interpolation to very small scales in the image plane. A dedicated comparison of our ABM to IPM~\citep{Mediavilla2011ApJ...741...42M}, interpolation method~\citep{2022arXiv220505141D}, and contour method~\citep{WyitheWebster1999Contours} is warranted, and is left for future work.

\subsection{General description of the ABM}
\label{ssec:overview}

Consider a stellar source of radius $\beta_{\rm S}$ (say, in angular units, for instance) moving 
behind a gravitational lens, along a predetermined path in the source plane, which we assume here to be a square region for illustration purposes. The length of this path thus determines the area in the source plane that has to be modeled. This area, along with the background convergence ($\kappa_{\rm m}$) and shear ($\gamma_{\rm m}$), define the relevant area in the image plane. Once the area in the image plane is determined, it is divided into $n_{\rm p}{\times} n_{\rm p}$ pixels and each pixel is mapped to the source-plane using the lens equation. Only those pixels in the image-plane that are mapped back to within $\Delta \beta$ from the source center ($\beta_1, \beta_2$), are secured for the next iteration. In the next iteration, each of the secured pixels is divided into new $n_{\rm p}{\times} n_{\rm p}$ subpixels to be mapped back to the source plane, while the source-plane boundary is also reduced by some
pre-specified factor $\eta$, i.e., $\Delta \beta \coloneqq \Delta \beta / \eta$, with $\eta>1$. This recursive process goes on until $\Delta \beta$ is equal to the source size. In the end, we are left with a list of all the high-resolution image-plane pixels that map back to the source (note the pixel size of each image-plane pixel is now also of the order of the source size). The ratio between total area of the (micro) images in the image-plane and the source area in the source-plane, gives the magnification of the source at this position. 

To better describe the method, let us define the source plane region of interest as a square of length ${\rm L} = v_{\rm t} {\times} t_{\rm tot}$, where $v_{\rm t}$ is the relative transverse source velocity in the source plane, and $t_{\rm tot}$ the total length of the light curve in time. For a given combination of macro-convergence ($\kappa_{\rm m}$) and macro-shear ($\gamma_{\rm m}$), the corresponding tangential and
radial magnification values are $\mu_{\rm t} = (1-\kappa_{\rm m} - \gamma_{\rm m})^{-1}$ and $\mu_{\rm r} = (1-\kappa_{\rm m} + \gamma_{\rm m})^{-1}$, respectively. In general, the macro-shear has two components, ($\gamma_{\rm 1,m}$, $\gamma_{\rm 2,m}$), but one can always find a coordinate frame in which  $\gamma_{\rm m} {=} \sqrt{\gamma^2_{\rm 1,m} {+} \gamma^2_{\rm 2,m}}$ and $\gamma_{\rm 2,m} {=} 0$. Hence, the image plane region of interest will be a rectangle of size $\theta_{\rm t} {\times} \theta_{\rm r}$, where $\theta_{t}= \epsilon({\rm L}/{\rm D}_{\rm s})\mu_{t}$ and $\theta_{r}= \epsilon({\rm L}/{\rm D}_{\rm s})\mu_{r}$, with $\epsilon(\geq {1})$ and ${\rm D}_{\rm s}$ being the angular diameter distance to the source. The $\epsilon$ factor will lead to a larger image-plane area considered. It is introduced to create a buffer in order to control missing rays: larger $\epsilon$ values lead to a smaller number of missing rays. We usually use a typical value of $\epsilon{\sim}4$ leading to ${<}1\%$ missing rays. This was verified by calculating the minimum number of stars that one has to take into account such that the fraction of diffusive flux due to a far away star, compared to the total flux, is less than ${<}1\%$ \citep{1987A&A...171...49S, WAMBSGANSS1999353}, and by making sure that we are well beyond that threshold. We have also confirmed this by simulating a larger box (${>}2{\rm L}$) with a lower resolution in the source plane and using some visual inspections similar to \citet{WAMBSGANSS1999353}. In general, the user should note that higher $\epsilon$ values also increase the required computational resources.

\begin{figure*}
    \centering
    \includegraphics[height=14cm, width=14cm]{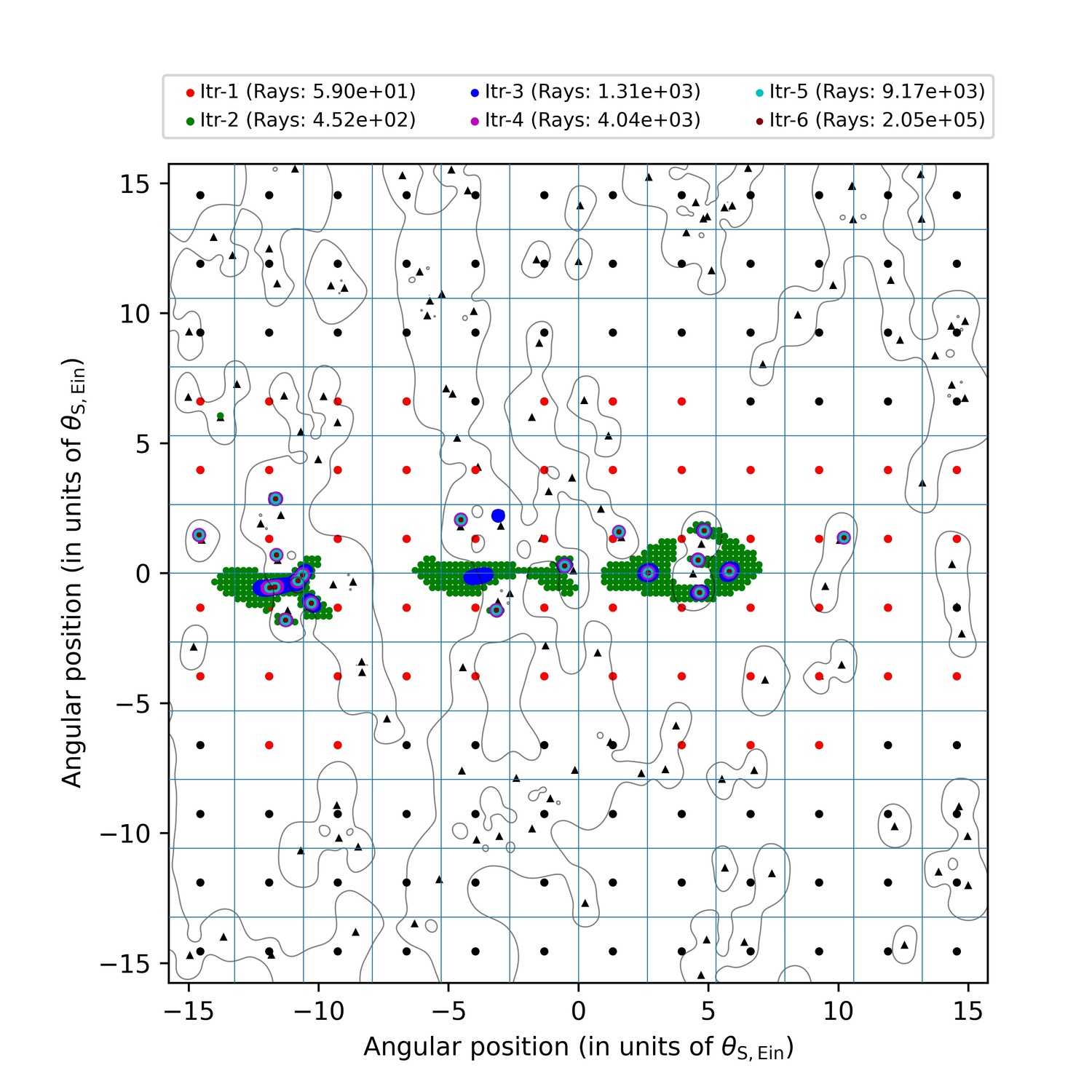}
    \caption{Example of an adaptive refinement of a lens plane in ABM. The lens plane is initially pixelated with $12{\times}12$ pixels (\emph{light-blue grid}). The positions of microlenses are shown by \emph{small black triangles}, and the critical curves are shown by \emph{thin black curves}. In the first iteration, the centers of all image-plane pixels (marked by \emph{black dots}) are shot to the source plane. Pixels that map to within $\Delta\beta$ from the source position are kept and re-marked here as \emph{red dots}. These pixels are those that are refined in the next iteration. Similarly, pixels identified as close enough to the source center in the 2nd, and 3rd iterations are shown by \emph{green} and \emph{blue dots}, respectively. Relevant pixels identified in 4th, 5th, and 6th iterations are shown as \emph{cyan}, \emph{light-green}, and \emph{maroon dots}, respectively. The number of accepted pixels (rays) from each iteration is indicated in the \emph{Legend}. The figure demonstrates how the algorithm iteratively converges onto the true position of the images as the image-plane resolution in the relevant pixels, and the source-plane boundary, are adaptively refined until $\Delta\beta$ equals the source radius. Axes are given in units of $\theta_{\rm s,Ein}$, where $\theta_{\rm s,Ein}=1.9151$ micro-arcseconds, so that each initial pixel in the light-blue grid is roughly $\simeq$5 micro-arcseconds (2.627${\times}\theta_{\rm s,Ein}$). The example lens plane presented is chosen to be a square for illustration purposes.}
    \label{fig:abm_refinement}
\end{figure*}

Although the square region with side L covers the specified source plane region, in order to be certain that no useful pixels in the image plane are missed, we start the simulation with an effective source-plane radius, $\beta_0$, that is a few times larger than L.

Given the specified source size, $\beta_{s}$, initial choice of source-plane boundary, $\beta_0$, and  source-plane division factor $\eta$, the number of iterations, i.e., divisions that the program will perform, is obtained by the following equation:
\begin{equation}
    \beta_{s} = \frac{\beta_0}{\eta^{{\rm N}-1}},
    \label{eq:initial-final}
\end{equation}

In general, the above equation yields a non-integer solution for N. The total number of iterations is the round-up of N, i.e., its \emph{ceiling} $\lceil {\rm N} \rceil$. In each iteration, the relevant image plane pixels are further divided each into $n_{\rm p}{\times}n_{\rm p}$ higher-resolution pixels, and the source plane boundary is reduced by a factor $\eta$. For the final iteration, $\eta$ is adjusted such that $\eta \coloneqq \eta^{{\rm N}^*}$, with ${\rm N}^*$ being the remainder defined by ${\rm N} = \lfloor {\rm N} \rfloor + {\rm N}^*$, and $\lfloor {\rm N} \rfloor$ marks the \emph{floor} of N. 

Once the iterative procedure is done, we are left with those high-resolution image-plane pixels that were mapped to within 
the radius $\simeq\beta_{s}$ around the source position $(\beta_1, \beta_2)$ in the source plane.
Then we calculate the magnification factor ($\mu$) for the source at $(\beta_1, \beta_2)$, as the total multiple-image pixel area in the image plane, over the source area in the source plane. This can be shown to be simply: 
\begin{equation}
    \begin{aligned} 
\mu & = \frac{n \: \theta_{r} \: \theta_{t}}{n^{2 {\lfloor {\rm N} \rfloor} + 2}_{\rm p} \pi \beta_{\rm S}^2}, 
        \label{eq:magnifiation}
    \end{aligned}
\end{equation}
where $n$ is the total number of relevant pixels in the lens plane at the highest resolution (i.e., from the final iteration) and all other symbols 
have their usual meaning as defined above.
The above equation does not take into account the varying intensity profile of the source. In order to do so, we can divide the (final) source plane region, which is essentially the area of the source, into multiple pixels, and bin the rays accordingly (i.e., count the rays in each pixel). This is easily done as the source-plane position of each ray is already known. Then, the magnification of the source can be easily calculated, as a weighted average over the source, using the given light profile as the weight function. 

An example of a lens-plane refinement in ABM, demonstrating the procedure, is shown in Figure~\ref{fig:abm_refinement}. For illustration purposes, we choose a square lens-plane region. In the general case, where we initially consider a square source-plane size as described above, the lens plane will be rectangular, with similarly rectangular pixels.

To create a light curve in time, the procedure described above is applied to a set of evenly spaced $n_{\rm steps}$ points along the given source path. However, with the increase in the time resolution of the light curve, the distance covered by the source between two successive points on the light-curve will decrease and the image formation for these two (successive) points will not differ significantly. Hence, for a light-curve with resolution of ${\sim}1$ day, for example, instead of repeating the full calculation for each point in time, we can divide the whole light-curve into smaller sub light-curves (of length $sub\_t_{\rm tot}$ each) and perform the calculation for the whole sub light-curve at a time. Doing so allows us to minimise unnecessary repeats of the deflection-angle calculation, so it reduces the computation time but consumes more memory compared to the individual point calculation. In our implementation of ABM, the user can either choose to perform the calculation for individual time points as above, divide the light curve into sub light-curves, or even calculate the whole light curve in one go. 

We generate here simulated ABM light curves in various configurations, in Sections \ref{sec:comparison} and \ref{sec:cc_lc}, where further details are given .

\begin{figure*}
    \centering
    \includegraphics[height=7cm, width=8cm]{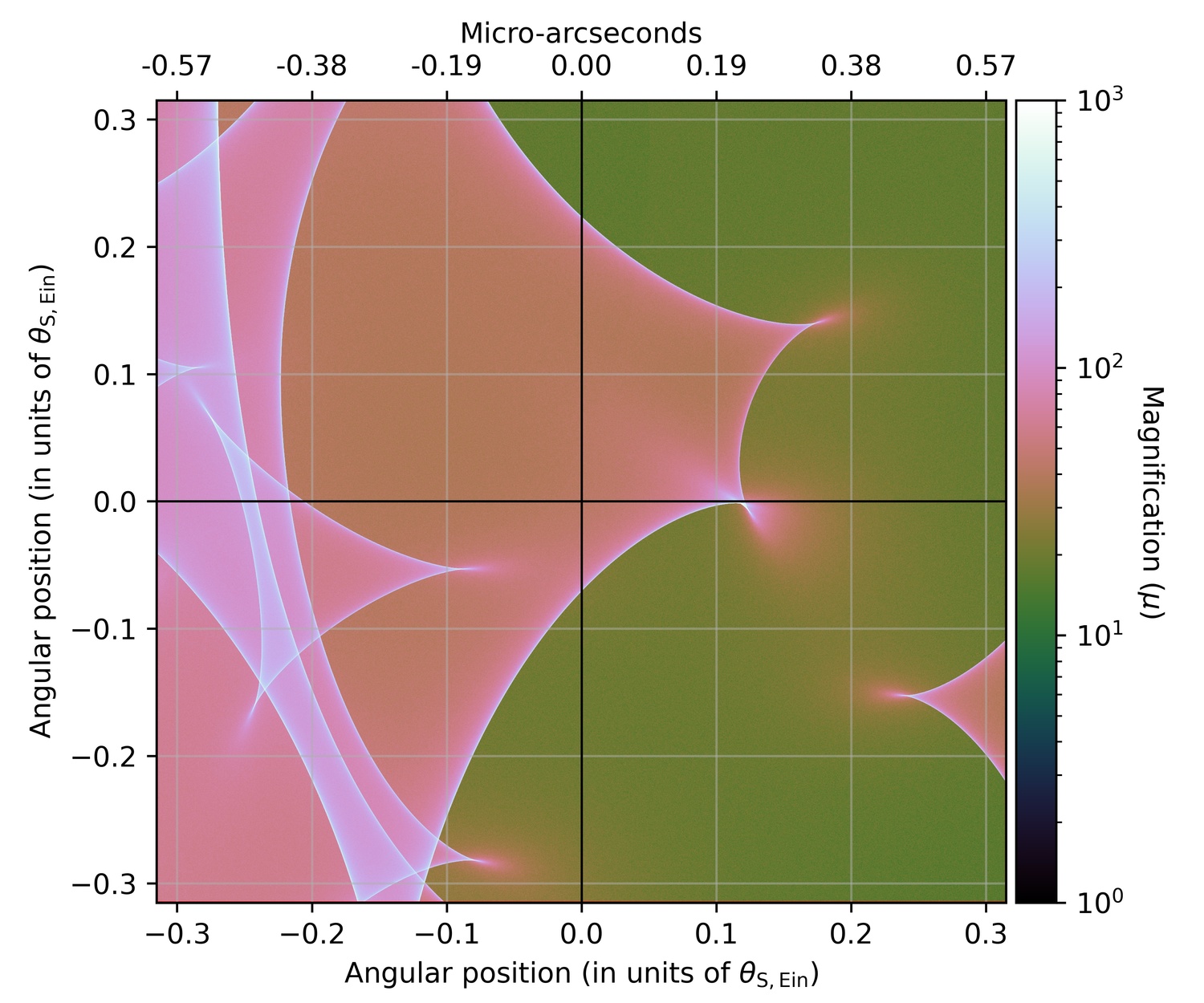}
    \includegraphics[height=7cm, width=8cm]{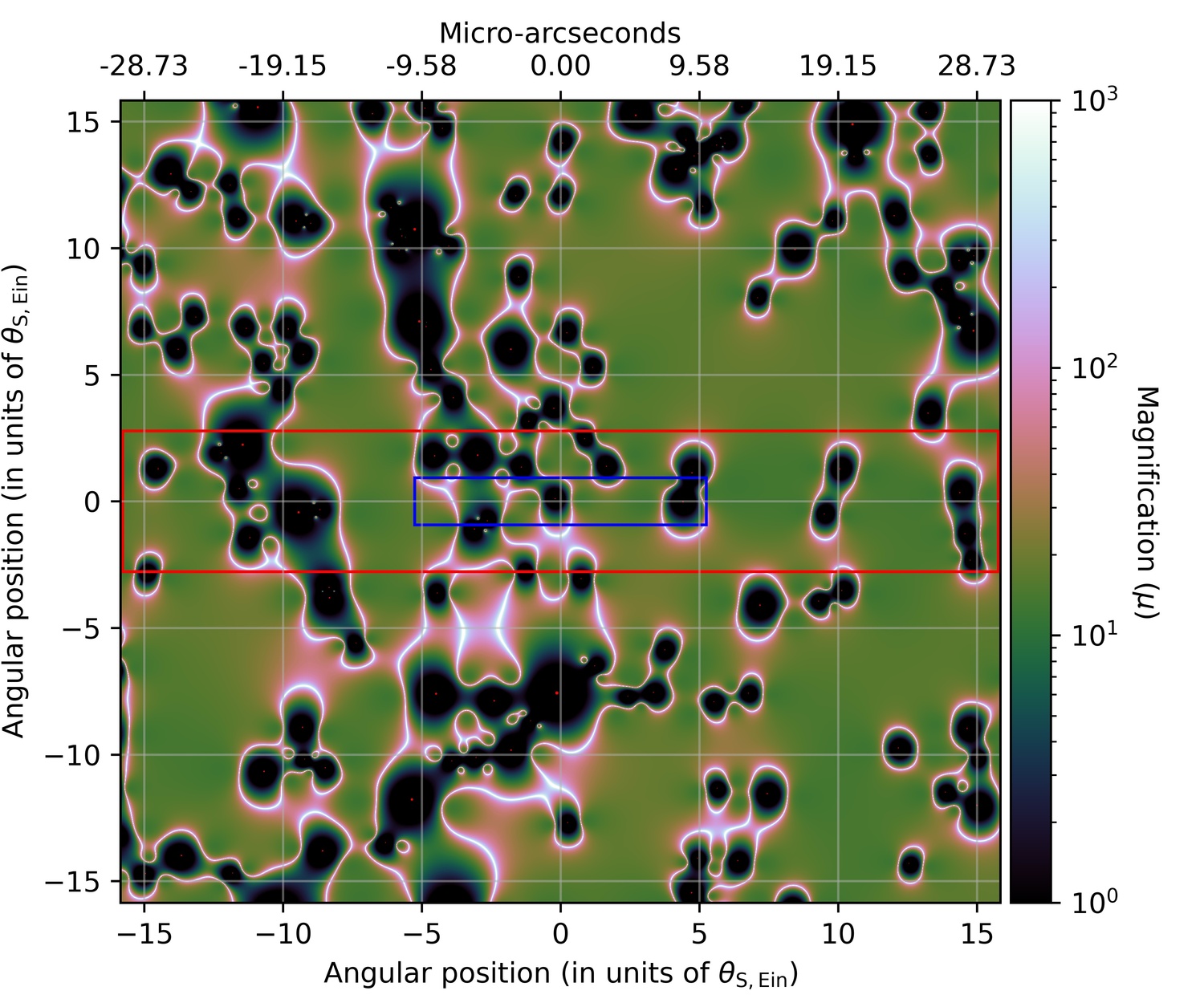}
    
    \vspace{0.5cm}
    
    \hspace*{-1.0cm}
    \includegraphics[height=6cm, width=16cm]{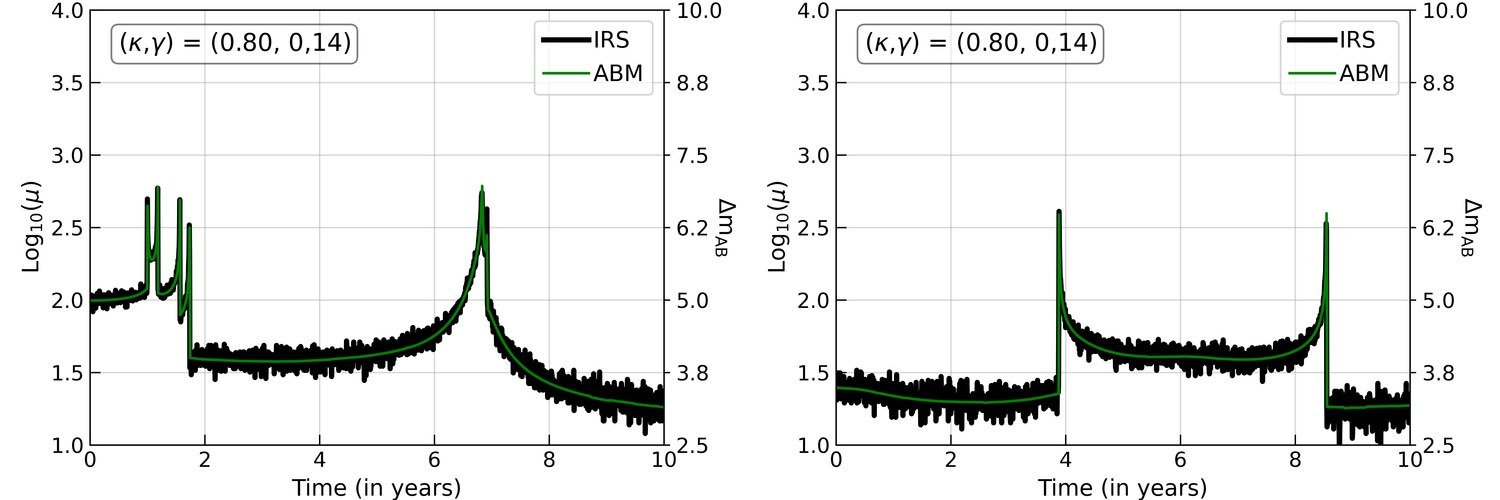}
    \caption{Example magnification map and resulting light-curves (corresponding to case-1 in Table~\ref{tab:tests}). The \emph{top-left} and \emph{top-right} panels show the magnification map in the source and image planes, respectively. Coordinates are given in units of the Einstein radius corresponding to 1${\rm M}_\odot$, for the chosen redshift configuration. The upper x-axis specifies the coordinates also in micro-arcseconds. The \emph{blue rectangle} in the \emph{right panel} represents the region in the lens plane corresponding to the source plane region seen in the left panel, as calculated using the macro-magnification values. The \emph{red rectangle} shows the region which has been used to shoot rays from the lens- to source-plane to construct the magnification map. The \emph{bottom-left} and \emph{bottom-right} panels show, respectively, the light-curve of a source with radius $10^{13}$ cm moving along the x and y axes in the source-plane map (\emph{black solid lines in top-left panel}). The \emph{black curve} and the \emph{green curve} represent the light-curves constructed using the IRS and ABM
    methods, respectively. The ABM method yields significantly less noisy and more accurate light-curves, in a significantly shorter time.
    } 
    \label{fig:cc_lc_0.80_0.14}
\end{figure*}

\begin{figure*}
    \centering
    \includegraphics[height=5.2cm, width=16cm]{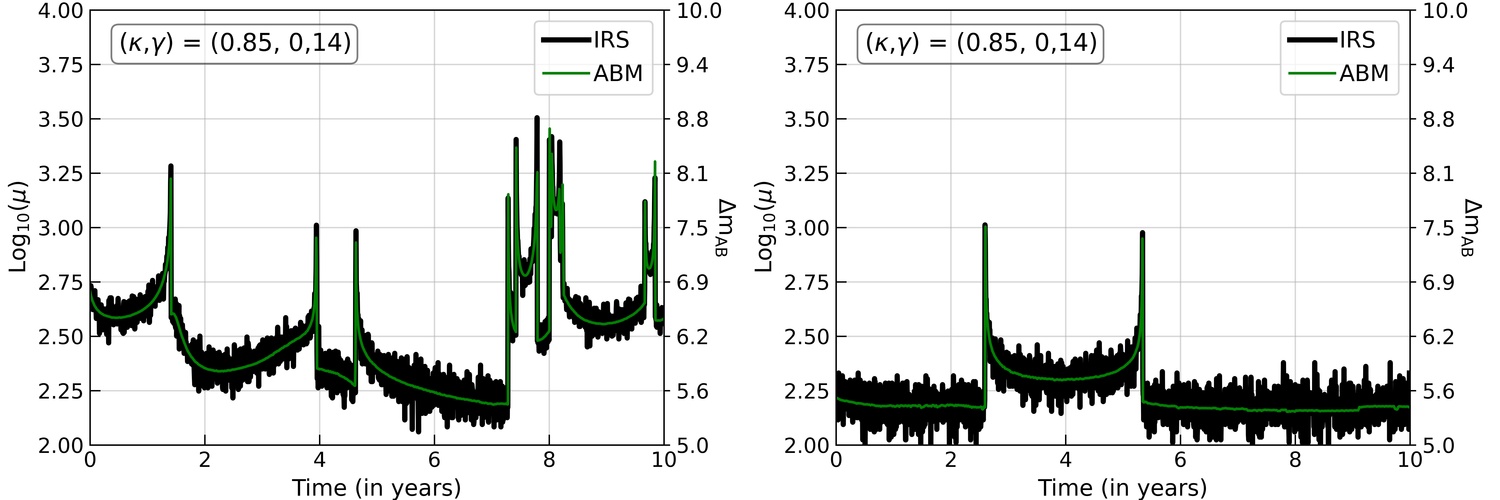}
    \includegraphics[height=5.2cm, width=16cm]{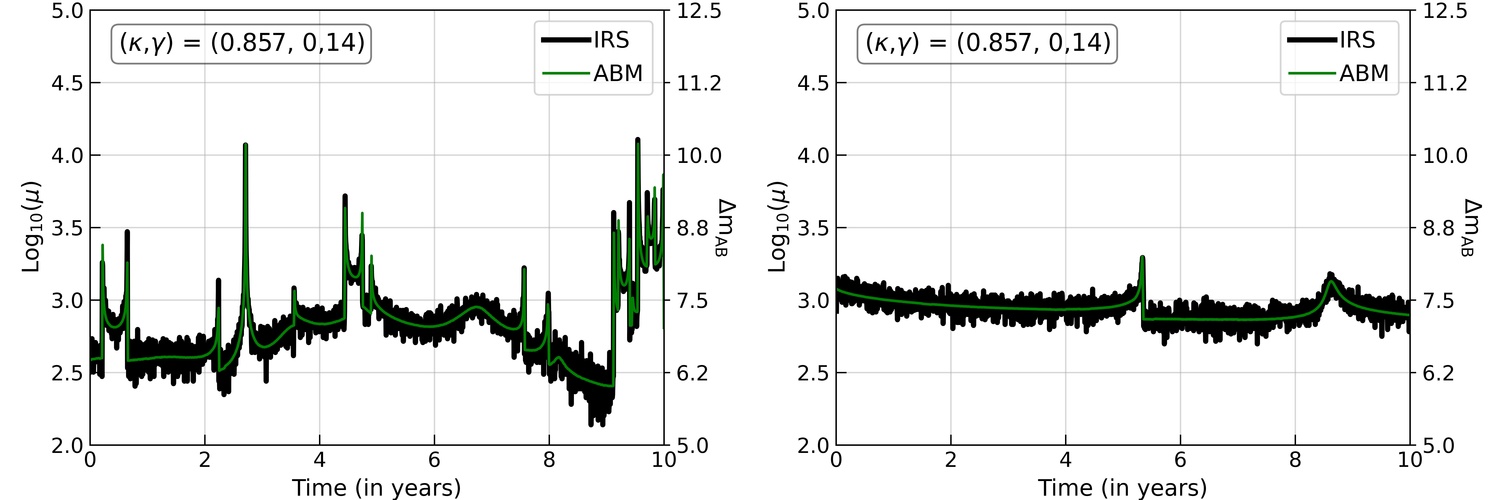}
    \includegraphics[height=5.2cm, width=16cm]{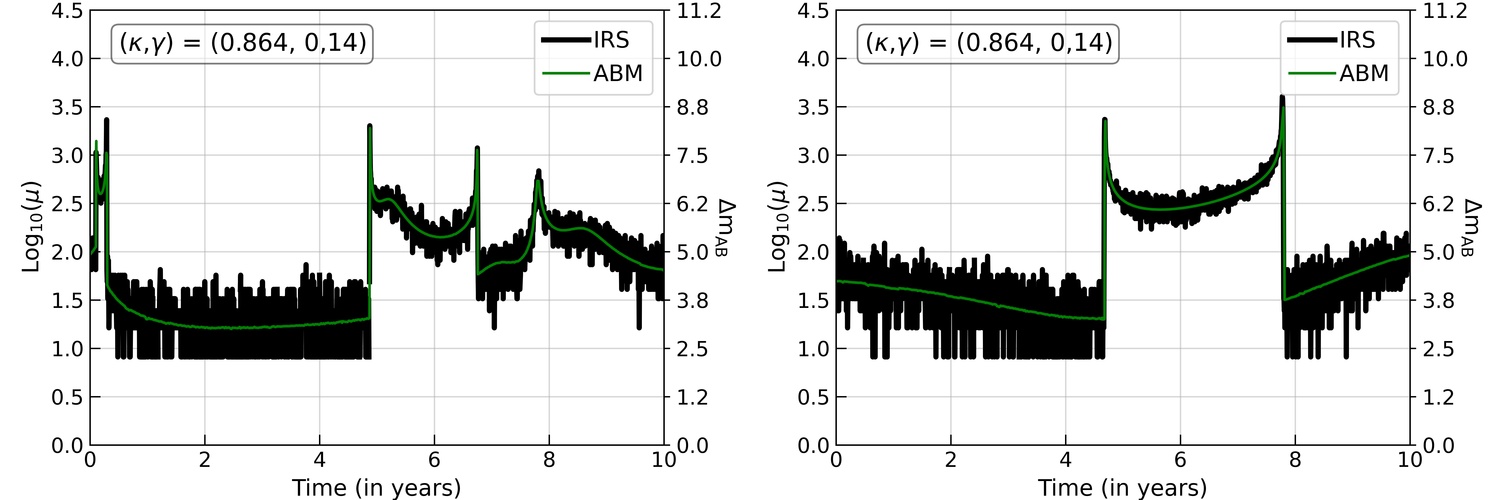}
    \includegraphics[height=5.2cm, width=16cm]{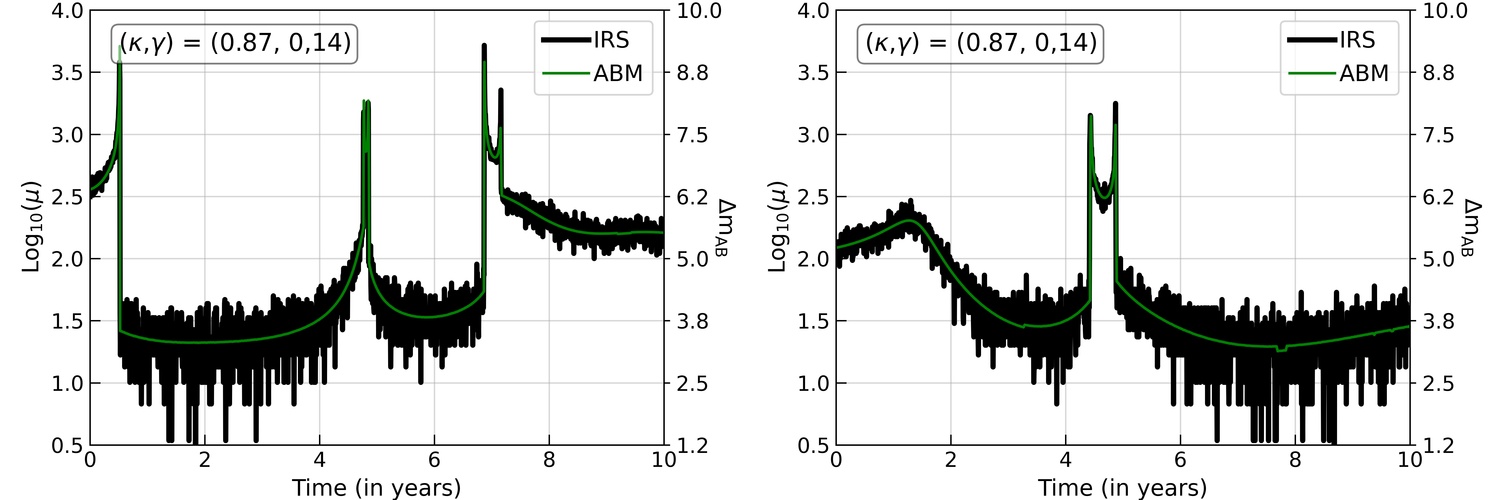}
    \caption{Light curves for different configurations from Table \ref{tab:tests} (see also Section~\ref{sec:comparison}). From top to bottom, the different rows show case-2 to case-5. For each case the \emph{left and right panels} show the light curves for a source moving parallel and perpendicular to the shear direction, respectively. The macro convergence and shear are also indicated for each plot. The black and green lines show, respectively,
    the IRS and ABM light curves for a $10^{13}$ cm source. The x-axis shows
    the observation time in years and the y-axis shows the corresponding magnification value, as well as the gain in apparent magnitude compared to the unlensed case. The comparison demonstrates that the ABM method works well in reproducing the true light curves, outperforming the IRS method in terms of both accuracy and efficiency.}
    \label{fig:cc_lc_test}
\end{figure*}

\begin{table*}
    \centering
    \caption{Different lensing configurations chosen for the comparison between ABM and IRS. \emph{Column 1:} case index; \emph{Column 2:} type of macroimage (minimum or saddle); \emph{Columns 3 \& 4:} the macro-convergence ($\kappa_{\rm m}$) and macro-shear ($\gamma_{\rm m}$) values; \emph{Column 5:} the fraction of macro-convergence ($\kappa_{\rm m}$) that is in the form microlenses; \emph{Column 6:} the corresponding surface mass density ($\Sigma_{\star}$) values; \emph{Column 7:} The smooth magnification corresponding to ($\kappa_{\rm m}{-}f_{\star}\kappa_{\rm m}$, $\gamma_{\rm m}$); \emph{Columns 8:} the macro-image magnification from ($\kappa_{\rm m}$, $\gamma_{\rm m}$); \emph{Column 9:} source sizes for which light curves are constructed; \emph{Column 10:} the average time taken to compute the source magnification for one point in the light curve. \emph{Column 11:} the total number of rays shot in ABM to generate the light curve. \emph{Column 12:} the (approximate) total number of pixels that are needed to be shot in standard IRS to achieve the same resolution as ABM. See text for further details.}
    \begin{tabular}{lcccccccccccr}
        \hline
        \# & Image type  & $\kappa_{\rm m}$ & $\gamma_{\rm m}$ &  $f_{\star}$ & 
        $\Sigma_{\star}$ &  $\mu_{\rm s}$ &  $\mu_{\rm m}$  &  Source size &  
        Average step time  & Rays shot & Rays to shoot\\
        &     &  &  &  & (${\rm M}_\odot/{\rm pc}^2$) &  &   &  (cm) & (seconds)  & in ABM & in IRS\\
        (1) & (2) & (3) & (4) & (5) & (6) & (7) & (8) & (9) & (10) & (11) & (12) \\
        \hline
        case-1 & Minimum & 0.80  & 0.14 & 0.10  & 186 & 17   & 49   & $10^{14}$  & ${<}1$      & $2.7{\times}10^7$  & ${\sim}10^{7}$ \\ 
               &        &       &      &       &     &      &      & $10^{13}$  & ${\sim}1$   & $2.8{\times}10^8$  & ${\sim}10^{9}$ \\
               &        &       &      &       &     &      &      & $10^{12}$  & ${\sim}2$   & $9.5{\times}10^8$  & ${\sim}10^{11}$ \\
               &        &       &      &       &     &      &      & $10^{11}$  & ${\sim}6$   & $2.3{\times}10^9$  & ${\sim}10^{13}$ \\
        \hline
        case-2 & Minimum & 0.85  & 0.14 & 0.01  &  20 & 181  & 345  & $10^{14}$  & ${<}1$      & $6.0{\times}10^6$  & ${\sim}10^{8}$ \\
               &        &       &      &       &     &      &      & $10^{13}$  & ${<}1$      & $5.5{\times}10^7$  & ${\sim}10^{10}$ \\
               &        &       &      &       &     &      &      & $10^{12}$  & ${\sim}2$   & $2.1{\times}10^8$  & ${\sim}10^{12}$ \\
               &        &       &      &       &     &      &      & $10^{11}$  & ${\sim}4$   & $6.0{\times}10^8$  & ${\sim}10^{14}$ \\
        \hline
        case-3 & Minimum & 0.857 & 0.14 & 0.01  &  20 & 296  & 1172 & $10^{14}$  & ${\sim}1$   & $1.1{\times}10^7$  & ${\sim}10^{9}$ \\
               &        &       &      &       &     &      &      & $10^{13}$  & ${\sim}13$  & $1.2{\times}10^8$  & ${\sim}10^{11}$ \\
               &        &       &      &       &     &      &      & $10^{12}$  & ${\sim}45$  & $4.1{\times}10^8$  & ${\sim}10^{13}$ \\
               &        &       &      &       &     &      &      & $10^{11}$  & ${\sim}110$ & $1.3{\times}10^9$  & ${\sim}10^{15}$ \\
        \hline
        case-4 & Saddle & 0.864 & 0.14 & -0.03 &  60 & -133 & -905 & $10^{14}$  & ${<}1$      & $1.4{\times}10^6$  & ${\sim}10^{9}$ \\
               &        &       &      &       &     &      &      & $10^{13}$  & ${\sim}2$   & $1.1{\times}10^7$  & ${\sim}10^{11}$ \\
               &        &       &      &       &     &      &      & $10^{12}$  & ${\sim}8$   & $6.5{\times}10^7$  & ${\sim}10^{13}$ \\
               &        &       &      &       &     &      &      & $10^{11}$  & ${\sim}47$  & $4.1{\times}10^8$  & ${\sim}10^{15}$ \\
        \hline
        case-5 & Saddle & 0.87  & 0.14 & -0.01 &  20 & -204 & -370 & $10^{14}$  & ${<}1$      & $2.7{\times}10^6$  & ${\sim}10^{8}$ \\
               &        &       &      &       &     &      &      & $10^{13}$  & ${\sim}1$   & $2.4{\times}10^7$  & ${\sim}10^{10}$ \\
               &        &       &      &       &     &      &      & $10^{12}$  & ${\sim}2$   & $1.0{\times}10^8$  & ${\sim}10^{12}$ \\
               &        &       &      &       &     &      &      & $10^{11}$  & ${\sim}3$   & $2.4{\times}10^8$  & ${\sim}10^{14}$ \\     
        \hline
    \end{tabular}
    \label{tab:tests}
\end{table*}

\subsection{Insight into the parameter space}
\label{ssec:parameters}
For running the program, in addition to the strong-lensing information (as well as redshifts, IMF limits, etc.), there are various physical parameters $(\beta_{s}, v_{\rm t}, t_{\rm tot}, f_*)$, and numerical parameters $(\epsilon_{draw}, \epsilon_{shoot}, \beta_0, n_{\rm p}, \eta, n_{\rm steps},sub\_t_{\rm tot})$, that the user has to specify to start the procedure. The details of these parameters are as follows: 

\begin{itemize}
    \item $\beta_{s}$: the radius of the stellar source (in cm). This is referred to by the program as the final, desired source-plane boundary.
    
   \item $\beta_0$: the initial source plane radius. The source plane radius in each iteration, $\Delta\beta$, is first set to the value of $\beta_0$, and then adaptively refined in each iteration, until reaching $\beta_{s}$. To make sure that no relevant rays are missed in the source-plane it is recommended to choose an initial source plane radius sufficiently larger than L (typically we adopt 4L for microlensing light curves near the main caustic, and 10L for ``on the main caustic'' cases, i.e., actual CCEs).

    \item $v_{\rm t}$: the transverse source velocity relative to the lens, in the source plane; i.e., relative to the caustics map. 
    In principle, $v_{\rm t}$ has contributions from both the bulk (or peculiar) velocity of the main lens compared to the observer, the relative velocity of microlenses in the lens, and the velocity of the source in the source plane. The typical bulk velocity of the main lens is ${\sim}500$ -- ${\sim}1000$ km/s and the microlenses within it can have velocities ${\sim} 1000$ km/s. However, when converting the typical microlens velocity from lens- to source-plane it becomes smaller compared to the bulk motion of the macrolens. Similarly, the source velocity contribution is estimated to be small due to the large distance to it. Hence, the relative source velocity, $v_{\rm t}$, can be said to be, approximately, the bulk velocity of the main lens \citep[e.g.,][]{Kayser1986, Oguri2017CausticCrossing}.
    In our work we set the relative source velocity ($v_{\rm t}$) to 1000 km/s. Other values can of course be specified, where in most cases light-curves can be rescaled to other choices of relative velocity, $v_{\rm t}'$ (in km/s), by multiplying the timeline ($x$-axes in the light curves) by a factor $v_{\rm t}'/1000$ \citep[][]{Diego2017CC}.
    
    \item $t_{\rm tot}$: the total duration of the light curve in the observer frame (in years).
    
    \item $f_*$: represents the fraction of macrolens convergence that is in microlenses, i.e., $f_* = \kappa_*/\kappa_{\rm m}$.
    
    \item $\epsilon_{draw}$ and $\epsilon_{shoot}$: factors determining the image-plane area in which microlenses will be distributed ($\epsilon_{draw}$), and the (smaller) image-plane area that will be pixelated and from which rays will be shot ($\epsilon_{shoot}$). These factors are introduced to control the number of missing rays in the simulation. Large values of $\epsilon$ generally imply 
    less missing rays; however, they also increase the computation time significantly. We find that typical values of $\epsilon_{draw}\sim10$, and $\epsilon_{shoot}\sim$3-4 are usually sufficient to obtain high accuracy with very little missing rays ($<1\%$). Note that $\epsilon$ values are only needed when running microlensing simulations in the moderate regime (where the convergence and shear are fixed). For actual CCEs, one specifies the desired region manually.

    \item $n_{\rm p}$: the number of subpixles on each axis by which each (valid) pixel will be divided in the next iteration. This is also the number of pixels on each axis in the first iteration. We find that a value of $n_{\rm p}=12$ works very well. For small values of $n_{\rm p} \: ({<} 10)$, it is required to choose larger $\beta_0$ to compensate for the small number of pixels in the image plane, allowing for more iterations (and potentially increasing computation time).
    
    \item $\eta$: the factor by which the initial source plane will be reduced in each iteration. Typically a value of $\eta=0.7n_{\rm p}$, we have found, works well.
    
    \item $n_{\rm steps}$: the number of evenly spaced points along the source path in which the light curve will be calculated. As the routine is quite flexible, one can easily change these numbers. We recommend to begin with a sufficiently high density of steps in order not to miss individual peaks. 
    
    \item $sub\_t_{\rm tot}$: the length of the ``sub light-curve" (in units of years) to which the main light curve is divided. This parameter is set to zero when one wants the code to calculate the magnification for one time-step at a time, and irrelevant when one wants the code to calculate the magnification for the whole light curve in one go.
    
\end{itemize}

\section{Microlensing Light Curves and Comparison with IRS}
\label{sec:comparison}

We now first demonstrate the performance of the ABM method by generating a variety of light curves for different scenarios, and comparing them to light curves from simple IRS (with a sufficiently large number of rays). In this section we work in the ``moderate"-magnification regime, where the source is placed ${\sim}10^{-3}$ -- ${\sim}10^{0}$ arcseconds from the main lens caustic, and the underlying magnification is typically of order $10^{3}$ -- $ 10^{1}$ (see Table~\ref{tab:tests}). We choose five combinations of the macrolens convergence ($\kappa_{\rm m}$) and shear ($\gamma_{\rm m}$) such that they represent macro images forming at minima and saddle-points of the time-delay surface, with different magnification values. We do not consider here maximum macro-images, as they usually form near the center of the lens and are de-magnified if far from the critical curve (however, for completeness, in \S~\ref{ssec:cc_ml_rad} we consider one CCE case where one image is maximum and the other is saddle).

We show here an observational time span of 10 years and adopt a relative source velocity of 1000 km/s. We also choose for each combination some value for the fraction of convergence that is in the form of point masses, determining the corresponding surface mass density in point masses or stars ($\Sigma_{\star}$). The latter
should be at least equivalent to the typical intra-cluster medium (ICM) microlens density, which is around 1-10 ${\rm M}_\odot/{\rm pc}^2$ (depending, for example, on the total mass density and distance from the cluster center). In principle one can increase the microlens density value arbitrarily, but doing so will also increase the computation time. 
Once these inputs are specified, $\kappa_{\star}$, i.e., the convergence corresponding to $\Sigma_{\star}$, is removed from $\kappa_{\rm m}$, to get the smooth matter convergence and magnification ($\mu_{\rm s}$). 
We adopt a Salpeter initial mass function \citep[IMF;][]{1955ApJ...121..161S} as the mass function for the microlenses, in the range [0.08, 1.5]~${\rm M_\odot}$. These limits follow from the fact that most of the ICM stars are old \citep[e.g.,][]{2016MNRAS.461..230E, Kelly2017CC}. The choice of exact mass function here is flexible, and could be easily updated or replaced.
To make sure that no rays were missed, we distribute microlenses in a large circular region of radius $4\theta_t$ (see Section~\ref{sec:abm} for more details about the parameters) and only pixelate a rectangle region with sides $6[\theta_t, \theta_r]$ within it.
The above-mentioned steps are similar for both minimum and saddle cases except for one difference: for a minimum image, removing the $\kappa_{\star}$ from $\kappa_{\rm m}$ results in a smaller smooth magnification ($\mu_{\rm s} {<} \mu_{\rm m}$) whereas for saddle it increases the smooth magnification ($\mu_{\rm s} {>} \mu_{\rm m}$).  Hence, for saddle cases, instead of removing $\kappa_{\star}$ from $\kappa_{\rm m}$, we add
it onto it, such that $\mu_{\rm s}$ always remains smaller than $\mu_{\rm m}$, as we are
calculating the lens plane area using $\mu_{\rm m}$. 
We stress that such a difference in the minimum and saddle image simulations, does not introduce any change to the overall light-curve simulation process and one can also simply calculate the image-plane region using the smooth magnification value ($\mu_{\rm s}$).

\begin{figure*}
    \centering
    \includegraphics[height=8cm, width=16cm]{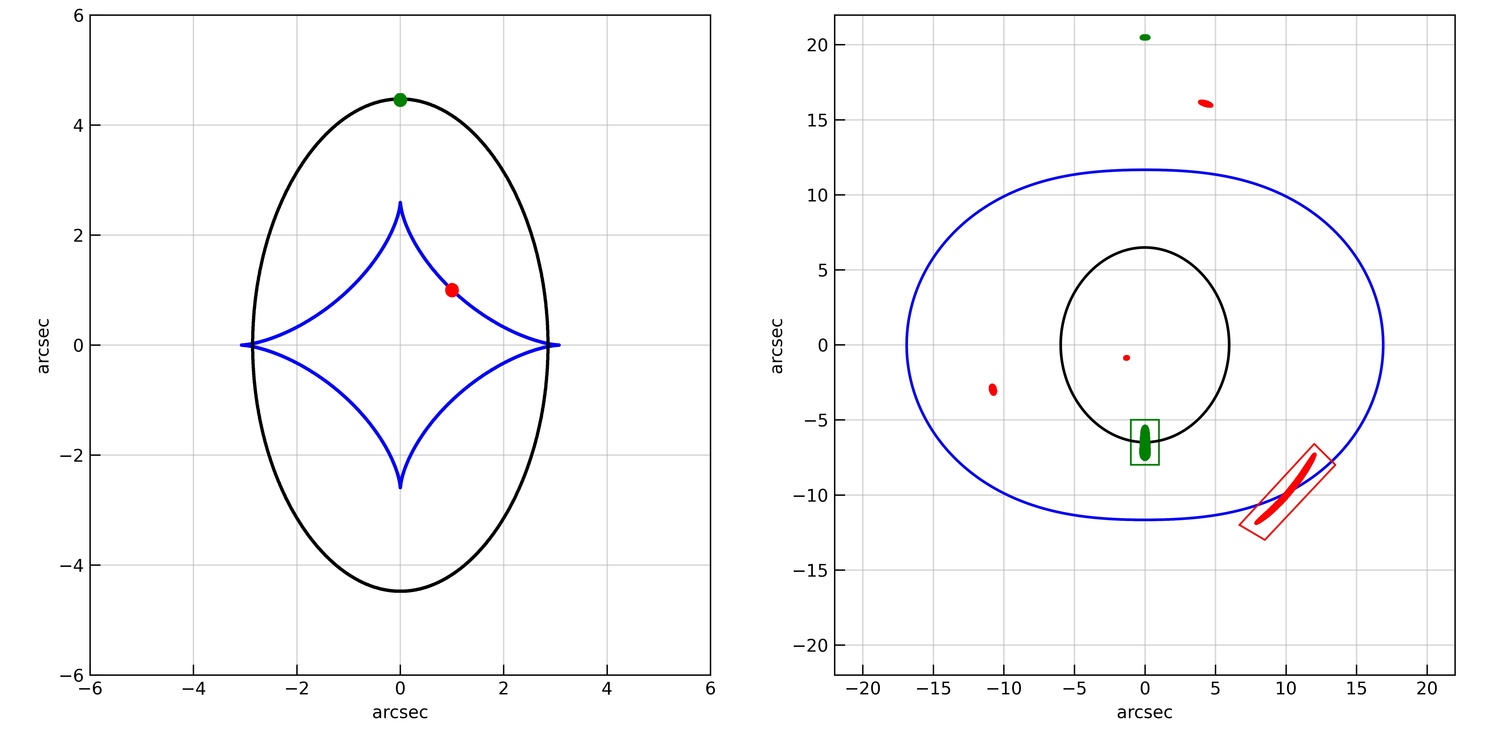}
    \caption{Caustics and critical curves of an input macrolens. The \emph{left and right panels} show the caustics and critical curves in the source and lens planes, respectively. The \emph{blue(black) curves} correspond to the tangential(radial) caustics and critical curves. The \emph{red and green dots} in the left panel represent two sources sitting on the tangential and radial caustics, respectively, and the corresponding images are shown in the right panel. The \emph{green and red boxes} mark the merging images on the radial and tangential critical curves, i.e., those relevant for CCEs.}
    \label{fig:macro_plot}
\end{figure*}

\begin{figure*}
    \centering
    \includegraphics[height=7cm, width=8cm]{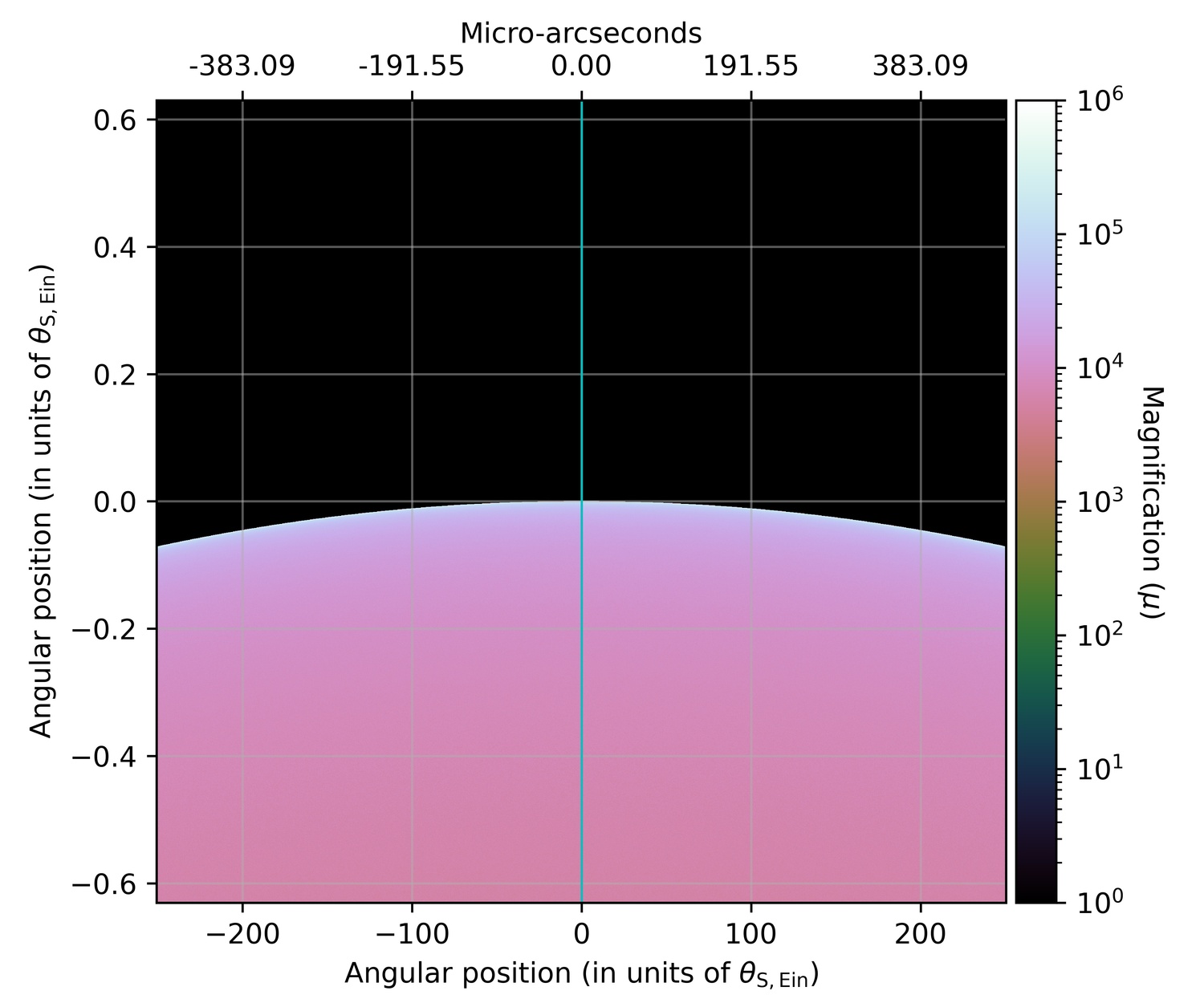}
    \includegraphics[height=7cm, width=8cm]{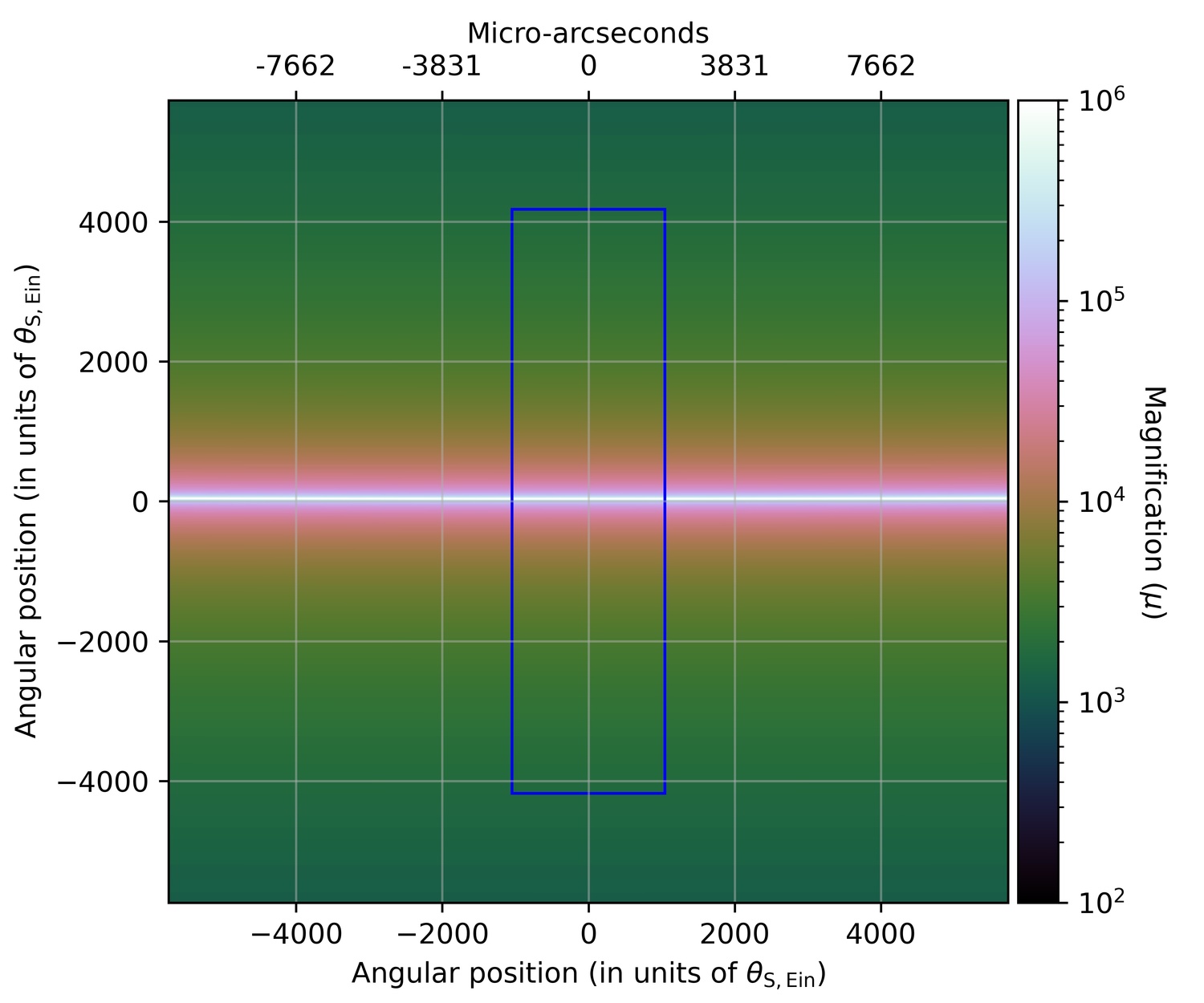}
    
    \vspace{0.5cm}
    
    \includegraphics[height=7cm, width=16cm]{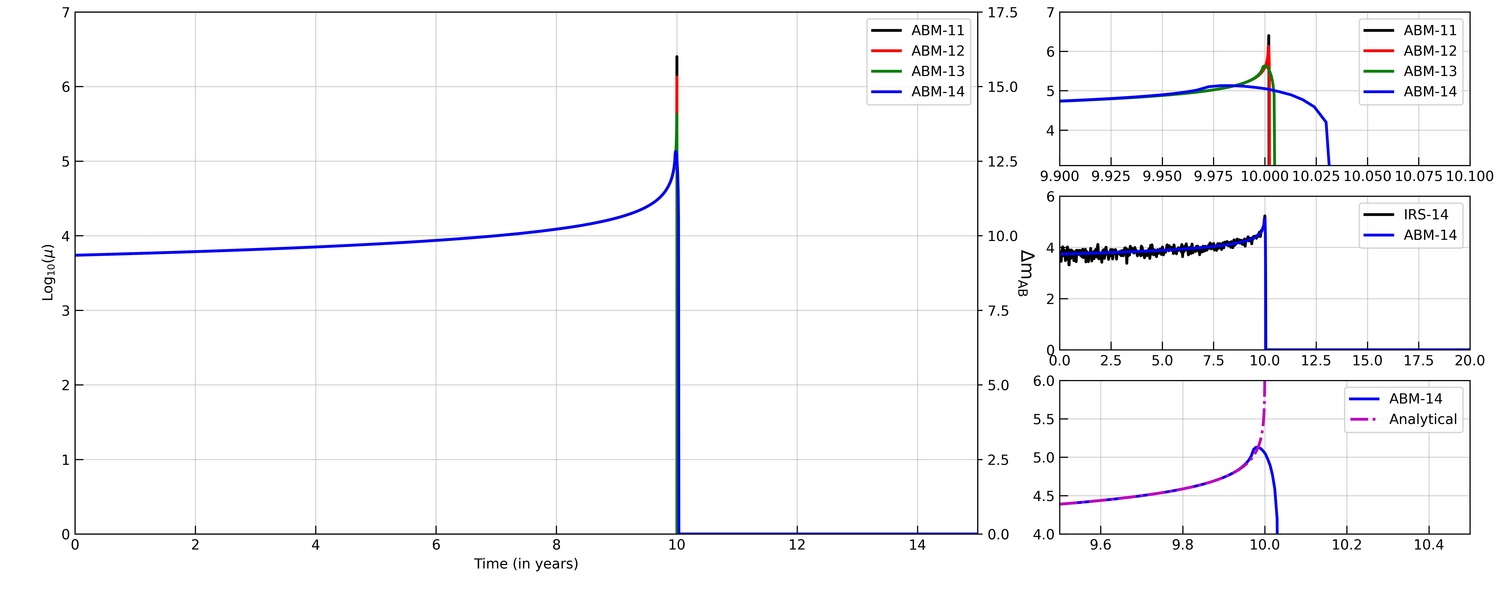}
    \caption{ABM light curve, in the absence of microlenses, for a radial CCE (the green source seen in Figure~\ref{fig:macro_plot}). The 
    \emph{top left} and \emph{top right} panels show the magnification map
    in the source and image planes, respectively.
    The blue box represents the region from which rays have been shot. The primary x- and y-axes are in units of Einstein radius of 1${\rm M}_\odot$ and the secondary x-axis is in
    units of micro-arcseconds. In the \emph{main bottom panel}, the blue, green, red, and black lines shows
    the light curve for a source with radius $10^{14}$, $10^{13}$, $10^{12}$, and $10^{11}$ cm,
    respectively. The x-axis represents
    the observational time in years and the y-axis shows the source magnification. The source crosses the 
    caustic at ten years. Prior to the caustic crossing the magnification for different size sources is similar.
    However, at the caustic crossing, different sources attain different magnification depending on
    their size: smaller sources get more highly magnified. After the crossing the magnification drops
    to zero. The process can also occur in the reverse direction. The \emph{upper inset plot} shows a zoomed-in view of the caustic crossing. The \emph{middle
    inset plot} shows the comparison of the ABM light curve (blue) with IRS light curve (black) for a 
    source with radius $10^{14}$ cm. In the bottom-right panel the ABM
    light curve is compared with the analytical dependence ($\propto 1/\sqrt{\Delta y}$).}
    \label{fig:cc_lc_no_ml}
\end{figure*}

The results for case-1, for example, are shown in Figure~\ref{fig:cc_lc_0.80_0.14}. We show the magnification map in both the source plane and image plane, and the resulting light curves along the specified paths, for a $10^{13}$ cm source. The ABM light curves are plotted together with IRS light curves, for comparison. The latter were generated using a simple IRS method with a $2000{\times}2000$ grid in both the lens and source planes, and shooting from each pixel in the lens plane a 1000 rays, on average. 

The ABM curves are accurate and recover all the peaks in the light-curve very well. The fluctuations seen in the IRS curve are essentially noise; they are a result of the finite number of rays that are shot, and the small source size. The amplitude of these fluctuations can be decreased by increasing the total number of rays shot from the image to source plane, or via assigning different fractions of area elements to different pixels in the source plane as is done, for example, in IPM (\citealt{Mediavilla2006IPM}). The ABM method does not suffer from these fluctuations, as effectively, only rays from areas that fall on or initially, close to the source, are being further kept for successive iterations, allowing to probe the source in high resolution in a much smaller total number of rays (but at the cost of concentrating on one light curve at a time). 

Light curves for other test cases are shown in Figure~\ref{fig:cc_lc_test}.
The left and right columns show the light curves for sources traveling parallel and perpendicular to the shear direction (along the x-axis and along y-axis, in our simulations), respectively.
Again, for all cases it is evident that ABM preforms extremely well -- not exhibiting noticeable noise fluctuations and fully recovering the peaks.
It can also be seen in Figure~\ref{fig:cc_lc_test} that the peak values for IRS and ABM do not agree always. This is a direct result of the fluctuations in the IRS light curves, which also affect the peak values so they can be higher or lower compared to the actual value. 

For test cases 1, 2, and 3, it is also noticed that the parallel light curves contain more micro-caustic crossings than the corresponding perpendicular light curves. For case-1 (where magnification is relatively low), this is in part because of the microlens distribution in the image plane; if we redistribute the microlenses again randomly, we may get more caustic crossings in the vertical light curve as well.
However, for case-2 and case-3, it is mostly due to the fact that as we increase the 
macro-magnification value, the critical curves in the image plane get stretched in one direction, which results in stretching of the caustics in the source plane. Hence, a source moving perpendicular to the stretching direction would cross more caustics whereas a source moving parallel to the stretching direction would naturally cross less caustics in the same amount of time.
The same is also true for the saddle test cases (case-4 and case-5), i.e., more caustic crossings will occur perpendicular to the stretching direction. In these saddle-point cases, however, one observes high de-magnification regions in the source plane so that the overall magnification can be lower compared to the minimum cases.

As shown in Table~\ref{tab:tests}, for each test case, we have constructed light curves with source sizes of $10^{14}$, $10^{13}$, $10^{12}$, and $10^{11}$ cm. Each light curve is made of a total of 1000 points yielding a resolution of ${\sim}3.6$ days. This time resolution is more than sufficient for not missing peaks given the $\mu\propto 1/\sqrt{\rm \Delta y}$ behaviour next to peaks, with $\Delta y$ being the distance from the caustic (if needed more points can be then added around the peaks to increase the sampling resolution). The average time to calculate the magnification for one point in the different test cases is given in Table~\ref{tab:tests}, where we also note the total number of rays that one needs to shoot in IRS to achieve the same resolution that ABM achieves in the last iteration in the image plane. We note that the time specified for 1 light-curve point in ABM is the average time, calculated using two light curves: one along the x-axis and other along the y-axis, both containing 1000 points. In general, different points take different times depending on the corresponding magnification. Typically, low magnification points take less time, as the corresponding number of relevant pixels in the image plane is smaller. Points near the peaks in the light curve (denoting the crossing of micro-caustics in source plane) take relatively more time, as the number of relevant pixels in the image plane is higher. Looking at Table~\ref{tab:tests}, one can see that smaller source sizes and higher magnifications both increase the computation time, as may be expected. Smaller source size means that the number of iterations to go from the initial source-plane boundary ($\beta_0$) to the final source size ($\beta_{\rm S}$) is higher, requiring additional time for each point. Higher magnification values mean that the size of the relevant area in the image plane is larger, entailing more point mass lenses. Hence, the deflection angle calculation for each ray will become more expansive increasing the average time per point in the light curve. At present our ABM routine has been implemented in \textsc{python} and all the tests shown in Table~\ref{tab:tests} were done on a 8-core iMac with a 4.2 GHz processor.
\smallskip

\begin{figure*}
    \centering
    \includegraphics[height=6cm, width=18cm]{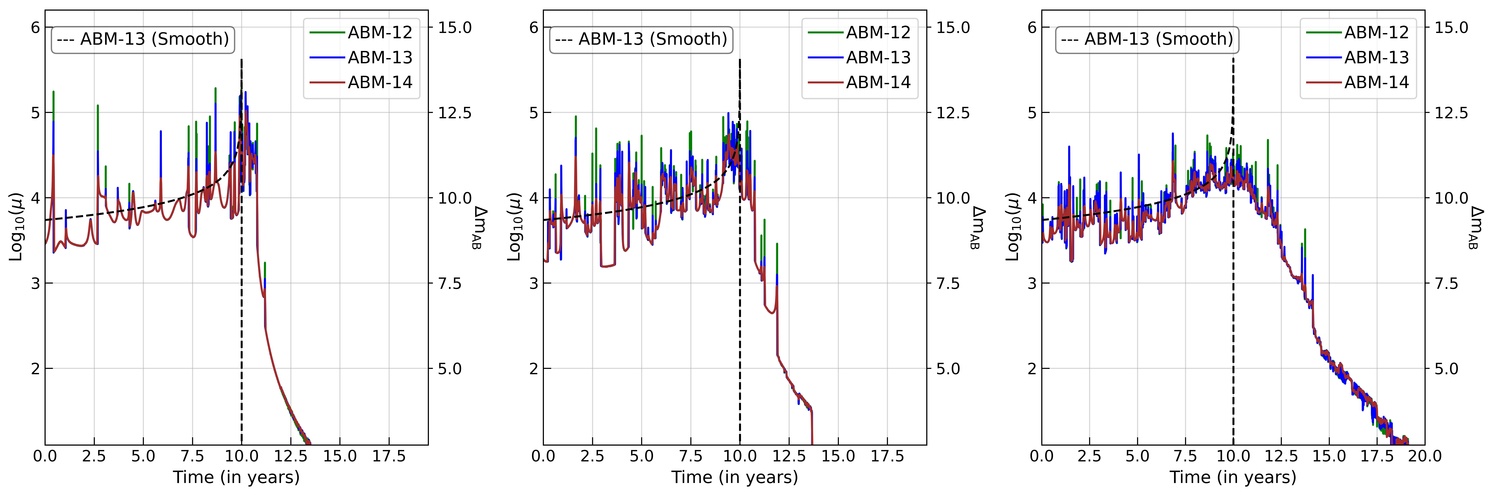}
    \caption{ABM CCE light curve in presence of microlenses for the green source (on radial caustic) in Figure~\ref{fig:macro_plot}: The left, middle, and right panels show the light curves for stellar densities of ${\sim}0.5$, ${\sim}2$, and ${\sim}8\:{\rm M}_\odot/{\rm pc}^2$. In all panels, the maroon, blue and green lines represent the light curves for a source with radius $10^{14}$, $10^{13}$, and $10^{12}$ cm, respectively. The x-axis represents the observation time and y axis shows the corresponding magnification value. Each peak in the light curve represents
    one micro-caustic crossing.}
    \label{fig:cc_lc_radial}
\end{figure*}

\section{Caustic Crossing Light Curves}
\label{sec:cc_lc}

In the previous section, for simulating the light curves we have assumed that the 
magnification is constant across the simulated region in the lens plane. 
Such an assumption is valid as long as the source (images) are not too close to the caustics (critical curves).
Once the source further approaches the caustic, the magnification grows as ${\propto} 1/\sqrt{\Delta y}$, where $\Delta y$ is the distance of the source from the caustic.
To show the applicability of ABM also in such scenarios, in this section we simulate light curves for sources very near -- essentially atop -- the main caustic, as they traverse it, i.e., CCEs.
To do this, we consider a simple strong lens system with a non-singular isothermal ellipsoid (NSIE) as the macro lens.
The velocity dispersion ($v_{\rm d}$) and ellipticity ($e$) of the lens are set to 1000 km/s and 0.1, respectively. The caustics and critical curves for such a system are shown in Figure~\ref{fig:macro_plot}.
One can see that the radial caustic is a fold caustic whereas the tangential caustic contains four cusps.
Observationally, most arcs in cluster lenses seem to consist of pairs of merging images -- mostly on the tangential critical curves, and sometimes on the radial one, so that the source usually lies on a fold caustic.
As a proof-of-concept, also here we concentrate on fold caustics. We examine both a radial and a tangential case (boxed images in Figure~\ref{fig:macro_plot}), where in the latter we place the source on the fold part of the tangential caustic, sufficiently far from the cusp points. 

We now detail the three CCE scenarios that we examine here: a smooth CCE where no microlneses are present; a realistic radial-caustic CCE (with microlenses); and a realistic tangential-caustic CCE (with microlenses).

\subsection{Caustic crossing without microlenses}
\label{ssec:cc_no_ml}

In the absence of microlenses (stars, black holes, etc.) in the macrolens, the critical curves and caustics form smooth, closed curves as shown in Figure~\ref{fig:macro_plot}.
The source magnification varies smoothly as we move away or towards the critical curve (typically of interest is the inner side of the caustic). This is demonstrated in Figure~\ref{fig:cc_lc_no_ml}, where we show the CCE light curve for a source crossing the radial caustic, i.e., the green source in Figure~\ref{fig:macro_plot}. In the smooth-lens case the caustic-crossing behavior is in principle universal and there is no fundamental difference between the radial and tangential cases. Hence, the radial case we show here is representative for both scenarios, whereas we just note that in the tangential case, the underlying magnification is a few percent higher.

To construct the corresponding source-plane magnification map, we shoot the rays from the blue rectangle. 
We divide the source plane and lens plane into $2000{\times}2000$ pixels and shoot on-average 1000 rays from each pixel in the lens plane. 
The y-axis of the source plane region is equivalent to a 20 year light-curve.
We increased the x-axis range so that we can also observe the curvature in
the radial caustic.
Doing so allows us to easily relate the left panels in Figure~\ref{fig:macro_plot} and \ref{fig:cc_lc_no_ml}. 
In the source plane (top-left panel), we see that in the positive part, above $y=0$, the magnification is zero. This is due to the fact that we are only considering the area near the merging images in the lens plane (green box in Figure~\ref{fig:macro_plot}).
That is, if we were to consider all the images in the lens plane, such a zero magnification region would not appear.
On the $y<0$ part of the map, the magnification drops as ${\propto} 1/\sqrt{\Delta y}$ as we move away from the caustic, where $\Delta y$ is the distance from the caustic \citep[e.g.,][]{1992grle.book.....S}.

\subsection{Caustic crossing with microlenses: Radial Image}
\label{ssec:cc_ml_rad}

The light curves generated in the absence of microlenses provide an excellent understanding of the underlying concept of CCEs. However, in practice, such a scenario is not observable since the lens contains a substantial amount of microlenses. As seen above, the presence of compact objects in the intracluster medium leads to the formation of additional critical curves in the lens plane around the microlenses. The size of these critical curves depend also on the background macro-magnification. For example, a microlens with mass M embedded in a macrolens with a background magnification of $\mu_{\rm m}$, will behave as a microlens with mass ${\sim}\sqrt{\mu_{\rm m}}$M \citep[e.g.,][]{Diego2017CC, Oguri2017CausticCrossing, Venumadhav2017}.

We now continue with the CCE light-curve simulations, but turn to the more realistic case and add microlenses near the merging image pair. In this subsection we focus on the radial merging image pair as well (green box in Figure~\ref{fig:macro_plot}). We adopt three different stellar density values for the microlenses: ${\sim}0.5~{\rm M}_\odot/{\rm pc^2}$, ${\sim}2~{\rm M}_\odot/{\rm pc^2}$ and  ${\sim}8~{\rm M}_\odot/{\rm pc^2}$. We remove the corresponding $\kappa_\star$ from the macrolens convergence ($\kappa_{\rm m}$) and center the lens plane region on the smooth critical curve.
Again we draw the microlenses according to a Salpeter IMF \citep{1955ApJ...121..161S} with a mass range [0.08, 1.5]~${\rm M}_\odot$ as before.
Near the radial critical curve, the radial magnification $\mu_{\rm r}$ varies significantly whereas the tangential magnification remains almost constant, which leads to the formation of images along the radial direction in the image plane with respect to the center of the lens.
We simulate here for the radial caustic case a rectangle patch of length 10.4 milli-arcsecond along the radial direction and 1.6 milli-arcsecond along the tangential direction, centered on the smooth critical curve. For the ${\sim}8~{\rm M}_\odot/{\rm pc^2}$ stellar density case, we considered a boxsize of 14.4 milli-arcsecond along the radial direction. These values were motivated by \citet{Diego2017CC} and are sufficient to cover the 20-year light curve. The size of the buffer region in this box is 0.2 milli-arcsecond. 

The light curves for the three microlens density values considered are shown in Figure~\ref{fig:cc_lc_radial}, for an observational time period of 20 years. The maroon, blue, and green lines show the light curve for a source with a radius of $10^{14}$, $10^{13}$, and $10^{12}$ cm, respectively. 
All light curves contain a total of 1000 points out of which 500 points uniformly cover the base light curve, and additional 500 points were distributed near the peaks, similar to the procedure described in Section~\ref{sec:comparison}. 
The average time taken for one point to be calculated in the case of ${\sim}0.5/2/8\:~{\rm M}_\odot/{\rm pc}^2$ is ${\sim}1/1/1$ seconds, ${\sim}1/2/9$ seconds, and ${\sim}3/9/37$ seconds for source radius $10^{14}$, $10^{13}$, and $10^{12}$ cm, respectively, on a 32-core system.

We can notice that, as we increase the microlens density, the frequency of peaks in the light curve increases, but the maximum magnification of these peaks decreases. The increase in the number of peaks can be understood from the fact that higher microlens densities introduce more point masses in the image plane which in turn increase the number of micro-caustic crossings in the source plane (note this is only true if the density of point masses is low enough; above a certain value there will be less peaks observed as more microelnses are added; see \citealt{2019A&A...625A..84D}). Apart from that, we also notice that when microlenses are present, the full caustic-crossing time gets longer compared to the crossing of a single smooth caustic. We observe peaks even 10 years past the nominal crossing time, as instead of a sharp macro-caustic there is a corrugated network of micro-caustics.

\subsection{Caustic crossing with microlenses: Tangential Image}
\label{ssec:cc_ml_tan}

A source lying on the tangential caustic gives rise to a merging image pair on the tangential critical curve in the image plane. The merging pair will be stretched in the tangential direction, with respect to the lens center, as shown by the red boxed image in Figure~\ref{fig:macro_plot} \footnote{In reality, many clusters exhibit elongated critical curves due to merging substructures in their centers; in such cases the merging image pair will be roughly perpendicular to the elongation direction of the critical curves}. For the NSIE lens model, the image formation near the tangential critical curve is not strictly perpendicular to it as was the case in the image formation near the radial critical curve (\S~\ref{ssec:cc_ml_rad}). Hence, to simulate the CCE light curves efficiently, we rotate the relevant image plane area such that the image formation is along the y-axis and rotate the source plane such that caustic lies along the x-axis\footnote{When running the code the user is given the option to specify rotation by a certain angle}. Then we simulate a rectangular box of size 1.2 milli-arcsec and 26.4 milli-arsec in the x and y directions, respectively. Again, we consider a buffer region of 0.2 milli-arcsec around the box to minimise the fraction of missing rays and pixelate the central region of size [0.8, 26] milli-arcsec. Compared to the radial case in \S~\ref{ssec:cc_ml_rad}, we have increased the size of the box along the y-axis, considering the fact that the tangential arc is more extended compared to the radial arc discussed above. Again we consider a Salpeter IMF in the mass range $[0.8, 1.5]~{\rm M}_\odot$ and simulate CCE light curves for three microlens density values, ${\sim}0.5, 2,$ and $8\:{\rm M}_\odot/{\rm pc}^2$.

\begin{figure*}
    \centering
    \includegraphics[height=6cm, width=18cm]{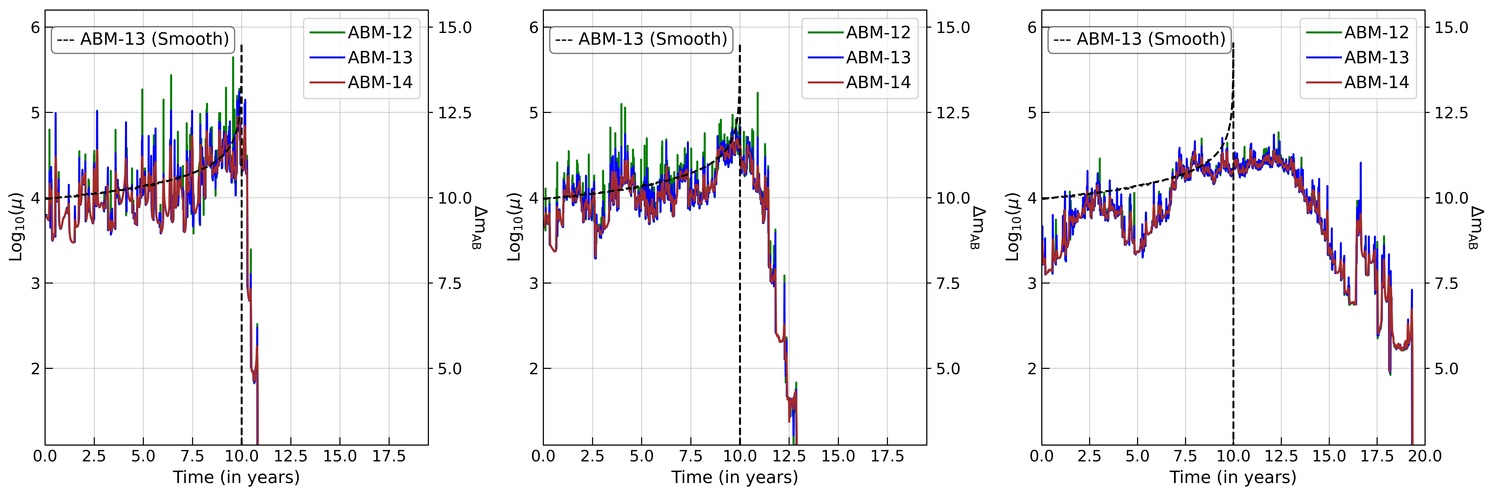}
    \caption{ABM CC light curve in presence of microlenses for the red source (on tangential caustic) in Figure~\ref{fig:macro_plot}: The left, middle, and right panels show the light curves for stellar densities of ${\sim}0.5$, ${\sim}2$, and ${\sim}8\:{\rm M}_\odot/{\rm pc}^2$. In all panels, the maroon, blue and green lines represent the light curves for a source with radius $10^{14}$, $10^{13}$, and $10^{12}$ cm, respectively. The x axis represents the observation time and y axis shows the corresponding magnification value. Each peak in the light curve represents one micro-caustic crossing. }
    \label{fig:cc_lc_tangential}
\end{figure*}

The corresponding light curves are shown in Figure~\ref{fig:cc_lc_tangential}. Also here we note that the number of peaks in the light curve increases with the microlens density, but the maximum magnification for individual peaks decreases. 
The average time taken for one point to be calculated with a point-mass density of ${\sim}0.5/2/8\:{\rm M}_\odot/{\rm pc}^2$ is ${\sim}1/1/3$ seconds, ${\sim}2/6/30$ seconds, and ${\sim}9/31/216$ seconds for source radii of $10^{14}$, $10^{13}$, and $10^{12}$ cm, respectively, on a 32-core system. As both Figure~\ref{fig:cc_lc_radial} and \ref{fig:cc_lc_tangential} represent caustic-crossing near a fold caustic, we observe a similar power-law behaviour $\propto1/\sqrt{\Delta y}$ (see Section \ref{ssec:cc_no_ml}). But as the proportionality constant depends on the underlying convergence value, the magnifications in the radial and tangential light curves differ by a constant ratio (see section 2 in \citealt{Venumadhav2017}). This can be clearly seen by the no-microlens cases shown by dashed lines in each panel of Figure~\ref{fig:cc_lc_radial} and \ref{fig:cc_lc_tangential}.

In all light curves shown in Figures~\ref{fig:cc_lc_radial} and \ref{fig:cc_lc_tangential}, we notice that not for all peaks in a light curve does the maximum magnification increase as we go down from a source size of $10^{14}$ cm to $10^{11}$ cm. This is because not all the peaks come from caustic crossing phenomena. Some peaks arise when the source passes near the cusp but does not actually cross it, leading to a somewhat wider peak in the light curve compared to a caustic crossing, albeit with somewhat lower magnifications. This stresses the importance of the ability to simulate light curves of small sources. Simulating instead light curves for larger sources (e.g., radius of $10^{14}$ cm), and then rescaling all the peaks analytically to imitate the magnification of smaller sources, as is sometimes proposed \citep[e.g.,][]{2019A&A...625A..84D}, should therefore be done with caution, although this approximation should indeed work well at least for the main peaks.\\

\section{Expected magnification for typical stars}
\label{sec:mu_stellar}

To put our light-curve simulations into observational context, we also include here an estimate of the minimal magnification values needed to observe different types of backgrounds stars (at the specified redshift of $1.5$, as an example). Comparing these values to the values seen in the light curves and the frequency of magnification light-curve peaks can give a rough idea of how many of these events we could potentially seek to observe. 

Different types of stars emit different amounts in various parts of the electromagnetic spectrum. Hence, the lens magnification needed to detect these different types of stars, for a given limiting magnitude, in a given observational band, can also differ substantially. 
We estimate here the required lens magnification to observe different kinds of stars in various \emph{HST} and \emph{JWST} filters, for several limiting magnitude depths.
The spectral energy distribution of different kinds of stars roughly follows a black body spectrum. 
The corresponding spectral radiance $B(\nu, T)$ is given as

\begin{equation}
    B(\nu,T) = \frac{2 {\rm h} \nu^3}{{\rm c}^2}\frac{1}{\exp(\frac{{\rm h} \nu}{{\rm k_B} T})-1},
    \label{eq:bb_radiance}
\end{equation}
where ${\rm h}$ is Planck's constant, ${\rm c}$ is the speed of light and ${\rm k_B}$ is 
Boltzmann's constant.
The corresponding specific luminosity $L_{\nu}$ is derived by integrating $B(\nu,T)$ over 
all solid angles and surface area of the source, given as

\begin{equation}
    L_{\nu}=4 \pi^2 R^2_{\rm s} B(\nu,T),
    \label{eq:bb_luminosity}
\end{equation}
where $R_{\rm s}$ is the radius of the source.
The observed flux density, $f_{\nu}(\nu_{\rm o})$, is related to rest-frame specific luminosity,
$L_{\nu}(\nu_{\rm e})$, through the differential form

\begin{equation}
    f_{\nu} \mbox{d}\nu_o=\frac{L_{\nu}}{4\pi D_{\rm L}^2}\mbox{d}\nu_e,
    \label{eq:flux_diff}
\end{equation}
where $\nu_{\rm o}$ and $\nu_{\rm e}$ are the observed and emitted frequencies respectively, 
and $D_{\rm L}$ the luminosity distance at redshift $z$. Further simplification gives

\begin{equation}
    f_{\nu}=\frac{L_{\nu}}{4\pi D_{\rm L}^2}(1+z).
    \label{eq:flux_luminosity}
\end{equation}

For a given filter with response function $Q_{\rm X}(\nu)$, the observed flux density
is \citep[][]{2002astro.ph.10394H},

\begin{equation}
    F_{\rm X} = \int \frac{f_{\nu}}{\nu_{\rm o}}Q_{\rm X}(\nu_{\rm o})\mbox{d}\nu_{\rm o}.
\end{equation}
Substituting the values from Equation \eqref{eq:bb_luminosity} and \eqref{eq:flux_luminosity} gives

\begin{equation}
    F_{\rm X} = \frac{\pi R^2_{\rm s}}{D_{\rm L}^2}(1+z) \int \frac{B(\nu_{\rm e},T)}{\nu_{\rm e}}
    Q_{\rm X}\left(\frac{\nu_{\rm e}}{1+z}\right)\mbox{d}\nu_{\rm e}.
    \label{eq:filter_flux}
\end{equation}

We use here the AB magnitude system \citep{1983ApJ...266..713O} and the
corresponding reference flux in filter $Q_{\rm X}(\nu)$ is given by
\begin{equation}
    F_{\rm X0}=\int \frac{10^{\frac{48.60}{-2.5}}}{\nu_{\rm o}}Q_{\rm X}(\nu_{\rm o})
    \mbox{d}\nu_{\rm o}.
    \label{eq:ref_flux}
\end{equation}
Hence, the change in the apparent magnitude of a source magnified by a factor $\mu$ is 
\begin{equation}
    m_{\rm AB}=-2.5\log_{10}(\mu F_{\rm X}/F_{\rm X0}).
    \label{eq:app_mag_lensed}
\end{equation}

In practice, various stellar atmospheric conditions modify the blackbody spectrum. For example, depending on its exciting and ionization state, the stellar Hydrogen can absorb much of the radiation below the Lyman or Balmer limits. Hence, in order to not underestimate the magnification that is needed to observe various stars, we use instead of a simple blackbody, updated \citet[][see also \citealt{CastelliKurucz2003IAUS..210P.A20C}]{castelli2004new} models downloaded from the STScI archive\footnote{\url{https://archive.stsci.edu/hlsps/reference-atlases/cdbs/grid/ck04models/ckp00/}}.  The results for different stars with a source redshift $z_{\rm s}=1.5$ are shown in Table 
\ref{tab:magnitude} for different HST filters, and in Table~\ref{tab:magnitude_jwst} for different JWST filters. 
One can see that different types of stars need different lens magnification thresholds in order
to be detected by the HST or JWST per given observational depth.
A Sun-like star needs a lens magnification in the range ${\sim}10^7 - {\sim}10^9$ to be detected with an observational depth of ${\sim}31 - {\sim}27$ AB mags, whereas a blue supergiant star only needs a lens magnification in the range ${\sim}10^2 - {\sim}10^4$ to be detected in the same magnitude range. Hence brighter stars are more likely to be detected, as indeed seems to be the case observationally \citep[e.g.][]{Kelly2017CC,Chen2019MACS0416CCE, Kaurov2019MACS0416CCE}.

Looking at our microlensing simulations it is evident that these minimal magnification values are frequently reached as stars move close the caustic, commensurate with the few observed lensed stars to date. There are a few competing effects in determining which stars should be easier to observe as caustic transients. For example, for main sequence stars, the luminosity grows much more quickly with mass than the radius, i.e., $L\propto{M}^{\alpha}$ with $\alpha\sim3.5-4$ and $R\propto M^{\beta}$,  with $\beta\sim0.5-1$, roughly. Although the magnification of smaller sources is generally, higher, growing roughly as $\sim1/\sqrt{R}$ \citep{Venumadhav2017}, the luminosity for higher mass stars is still sufficiently, differentially higher, explaining the strong bias towards brighter, massive stars. In other words, it seems that with current instruments one could mostly detect high-mass stars, and thus the currently relevant microlensing and CCE simulations for stars could mainly concentrate on $\sim10-100$ solar radii sources. 

While here we work with a fixed source redshift for simplicity, it is also worth noting that due to the blackbody-like spectral shape of stars, which -- following Wien's law -- peaks in a wavelength inversely proportional to the temperature, the peak flux for bright, massive stars falls bluewards of the typical red bands used in HST and JWST. This means that there is in fact a window for observing also high-redshift stars in lensed galaxies: While flux essentially drops as the luminosity distance squared, the blackbody spectrum shape means that for massive high-$z$ stars more flux will be redshifted to the observational band we use in practice, which would largely compensate for the drop in flux due to the larger luminosity distance. If we repeat the calculations as above, placing now the stars at redshifts 3, 6 and 9, we obtain that while the square of the luminosity distances suggest a flux decrease of $\sim5$, $\sim30$, $\sim70$, respectively (or, $\sim3$, $\sim10$, $\sim20$, if taking into account also the $(1+z)$ factor in eq. \ref{eq:filter_flux}), compared to $z=1.5$, the minimum required magnifications are only a few times higher, at most, than the current values in Tables \ref{tab:magnitude} \& \ref{tab:magnitude_jwst}. Therefore, brighter-type stars such as O and B stars that are highly magnified near caustics, should still be reachable with current facilities, namely \emph{Hubble} and \emph{JWST}, also at very high redshifts (basically, up to redshifts where the Lyman break falls into the observational band of interest).

\section{Conclusions}
\label{sec:conclusions}

In this work we have presented ABM - a simple and efficient method to simulate CCEs and nearby microlensing fluctuations for small sources. The method is based on the adaptive refinement of the relevant regions in the source and image planes, increasing the effective resolution. By focusing only on the relevant pixels in the image plane we show that one can significantly reduce the total number of rays that need to be shot from image to source plane for the magnification calculation. This leads to an increase in the both the accuracy and efficiency of the light curve calculation compared to simple IRS. We demonstrate the performance of ABM by simulating various light curves. First we generate light curves placing sources in a region of high magnification but far enough from the caustic so that the macro-lens effect can be taken into account as a constant macro-convergence and macro-shear. Then we simulated actual CCE light curves considering a simple macro-lens model with sources lying on the tangential and radial caustics, and with various values of microlens surface mass-density for both radial and tangential CCE cases. Both these options (constant convergence and shear, or an analytic macro model) are available in the publicly distributed code. An option for the users to feed their own existing lens models, is also in the works.

One drawback of the method is that at a time we can only simulate one light curve. However, as ABM takes significantly less time compared to IRS one can still simulate multiple light curves in a reasonable amount of time. We have shown that because the stellar luminosity grows faster with mass than stellar radius determining the magnification, there is a bias towards observing massive and giant stars (the magnification required to see them is lower), and that these stars could essentially be observed with typical red \emph{HST} and \emph{JWST} bands out to very high redshifts, because the $K$-correction for these compensates for the larger luminosity distance. For these star types, a typical light-curve time step in full-resolution ABM can range from about one second to a few dozen seconds, depending mostly on the underlying magnification, microlens fraction, source size, and number of computing cores available. Given the very high magnification involved in CCEs, such a procedure is essential for calculating the magnification in a sufficiently high resolution. We also note that the iterative refinement in ABM only focuses on the useful pixels and discards the useless pixels in the process. Doing so also allows us to efficiently use the available memory which is also important to consider when we simulate CCE light-curves on a work-oriented personal computer or laptop.

It should be highlighted also that in the current implementation of ABM there are opportunities to further increase the efficiency. For example, at present, the deflection angle calculation in ABM is exact. Hence, as the number of stars increase the deflection-angle calculation for each ray becomes more time consuming. We can make this part more efficient by calculating the deflection angle using tree structure, as discussed in \citet{WAMBSGANSS1999353}, for example. Another possible way (as mentioned above) to increase the efficiency of the ABM is to identify microlensing peaks in the light curve for a larger source and then use analytical approximation to construct the light curve for smaller sources (see e.g., \citealt{2019A&A...625A..84D}). This would allow the user to simulate approximate light curves in a relatively short time. One other caveat of the method is that due to the initial pixel size, relevant rays near microlenses may be missed. While the effect is evidently very small, this can be improved at a relatively low computational cost by adding smaller adaptive grids near each microlenses to recover the missing flux from micro-images. Such improvements, as well as the examination of these approximations, are planned for future work.

CCEs of individual small sources at cosmological distances constitute an exciting new field. These extreme magnification events allow us to observe small sources such as lensed stars at distances at which they would be far too faint to be observed otherwise. In addition, the microlensing fluctuations around caustics can help constrain the microlens mass-function of the intra-cluster medium (ICM), and thus allow for insight on both the ICM stellar population and the fraction -- and potentially mass-function -- of compact dark matter. To that end a comparison with simulated light-curves would be needed, which motivated us to formulate and publish the ABM code presented here. For meaningful conclusions one would need a considerable number of observations and monitoring of CCEs, some are expected soon from dedicated campaigns with the \emph{HST} and from the \emph{JWST}.

\section{Acknowledgements}
This work was supported by the Ministry of Science and Technology, Israel. We would like to thank Jos\'e Diego for insightful comments on the manuscript, and Erik Zackrisson, Tom Broadhurst, Pat Kelly, Liliya Williams and members of the Flashlight collaboration for useful discussions. The authors thank the anonymous referee of this work for useful comments. AZ also acknowledges support by Grant No. 2020750 from the United States-Israel Binational Science Foundation (BSF) and Grant No. 2109066 from the United States National Science Foundation (NSF).
This research has made use of NASA’s Astrophysics Data System Bibliographic Service.
Authors acknowledge the use of high performance computing facility Pegasus at IUCAA, Pune.
The colour scheme in the magnification maps is taken from \citet{cubehelix}.

\section{Data Availability}
The ABM \textsc{python} code is available at: \url{https://github.com/akmeena766/gl_cce}

\bibliography{reference}
\bibliographystyle{mnras}

\begin{table*}
    \centering
    \caption{Required magnification for different types of stars to be observe with HST in various
    bands and for given depths.}
    \begin{tabular}{lllcccccccc}
        \hline
        Spectral type & Mass (${\rm M}_{\odot}$) & Size (${\rm R}_{\odot}$) & Luminosity (${\rm L}_{\odot}$) & Temperature (K) & & $\mu_{\rm c}$ & & Comments \\
        & & & & & 27 AB & 29 AB & \textcolor{gray}{31 AB} & \\
        (1) & (2) & (3) & (4)&(5) & (6)& (7)&\textcolor{gray}{(8)}&(9)\\
        
        \hline
        &&&&F814W filter\\
        \hline
        
        G5V   & 1   & 1   &        1.00        & $5.750{\times}10^3$   & $4.4{\times}10^9$   & $7.0{\times}10^8$   & \textcolor{gray}{$1.1{\times}10^8$}   & Sun-like; Main-sequence\\
        B0V   & 15  & 6.5 & $2.04{\times}10^4$ & $3.000{\times}10^4$   & $1.3{\times}10^5$   & $2.1{\times}10^4$   & \textcolor{gray}{$3.4{\times}10^3$}   & Tau Scorpii; Main-sequence\\
        O3V   & 100 & 12  & $8.00{\times}10^5$ & $4.500{\times}10^4$   & $1.9{\times}10^4$   & $2.9{\times}10^3$   & \textcolor{gray}{$4.7{\times}10^2$}   & Main-sequence\\
        K5III & 1.2 & 45  & $4.40{\times}10^2$ & $4.000{\times}10^3$   & $3.6{\times}10^8$   & $5.8{\times}10^7$   & \textcolor{gray}{$9.1{\times}10^6$}   & Aldebaran; Red giant\\
        B8V   & 21  & 79  & $1.20{\times}10^5$ & $1.200{\times}10^4$   & $1.6{\times}10^4$   & $2.5{\times}10^3$   & \textcolor{gray}{$3.9{\times}10^2$}   & Rigel; Blue Supergiant\\
        
        \hline
        &&&&F105W filter\\
        \hline
        
        G5V   & 1   & 1   &        1.00        & $5.750{\times}10^3$   & $1.2{\times}10^9$   & $2.0{\times}10^8$   & \textcolor{gray}{$3.1{\times}10^7$}   & Sun-like; Main-sequence\\
        B0V   & 15  & 6.5 & $2.04{\times}10^4$ & $3.000{\times}10^4$   & $1.8{\times}10^5$   & $2.9{\times}10^4$   & \textcolor{gray}{$4.6{\times}10^3$}   & Tau Scorpii; Main-sequence\\
        O3V   & 100 & 12  & $8.00{\times}10^5$ & $4.500{\times}10^4$   & $2.9{\times}10^4$   & $4.5{\times}10^3$   & \textcolor{gray}{$7.2{\times}10^2$}   & Main-sequence\\
        K5III & 1.2 & 45  & $4.40{\times}10^2$ & $4.000{\times}10^3$   & $2.4{\times}10^7$   & $2.2{\times}10^6$   & \textcolor{gray}{$3.6{\times}10^5$}   & Aldebaran; Red giant\\
        B8V   & 21  & 79  & $1.20{\times}10^5$ & $1.200{\times}10^4$   & $8.0{\times}10^3$   & $1.3{\times}10^3$   & \textcolor{gray}{$2.0{\times}10^2$}   & Rigel; Blue Supergiant\\
        
        \hline
        &&&&F125W filter\\
        \hline
        
        G5V   & 1   & 1   &        1.00        & $5.750{\times}10^3$   & $7.4{\times}10^8$   & $1.2{\times}10^8$   & \textcolor{gray}{$1.9{\times}10^7$}   & Sun-like; Main-sequence\\
        B0V   & 15  & 6.5 & $2.04{\times}10^4$ & $3.000{\times}10^4$   & $2.3{\times}10^5$   & $3.7{\times}10^4$   & \textcolor{gray}{$5.8{\times}10^3$}   & Tau Scorpii; Main-sequence\\
        O3V   & 100 & 12  & $8.00{\times}10^5$ & $4.500{\times}10^4$   & $3.7{\times}10^4$   & $5.9{\times}10^3$   & \textcolor{gray}{$9.4{\times}10^2$}   & Main-sequence\\
        K5III & 1.2 & 45  & $4.40{\times}10^2$ & $4.000{\times}10^3$   & $4.6{\times}10^6$   & $7.3{\times}10^5$   & \textcolor{gray}{$1.2{\times}10^5$}   & Aldebaran; Red giant\\
        B8V   & 21  & 79  & $1.20{\times}10^5$ & $1.200{\times}10^4$   & $8.3{\times}10^3$   & $1.3{\times}10^3$   & \textcolor{gray}{$2.1{\times}10^2$}   & Rigel; Blue Supergiant\\
        
        \hline
        &&&&F140W filter\\
        \hline
        
        G5V   & 1   & 1   &        1.00        & $5.750{\times}10^3$   & $6.3{\times}10^8$   & $1.0{\times}10^8$   & \textcolor{gray}{$1.6{\times}10^7$}   & Sun-like; Main-sequence\\
        B0V   & 15  & 6.5 & $2.04{\times}10^4$ & $3.000{\times}10^4$   & $2.8{\times}10^5$   & $4.4{\times}10^4$   & \textcolor{gray}{$7.0{\times}10^3$}   & Tau Scorpii; Main-sequence\\
        O3V   & 100 & 12  & $8.00{\times}10^5$ & $4.500{\times}10^4$   & $4.5{\times}10^4$   & $7.2{\times}10^3$   & \textcolor{gray}{$1.1{\times}10^3$}   & Main-sequence\\
        K5III & 1.2 & 45  & $4.40{\times}10^2$ & $4.000{\times}10^3$   & $2.9{\times}10^6$   & $4.5{\times}10^5$   & \textcolor{gray}{$7.2{\times}10^4$}   & Aldebaran; Red giant\\
        B8V   & 21  & 79  & $1.20{\times}10^5$ & $1.200{\times}10^4$   & $9.4{\times}10^3$   & $1.5{\times}10^3$   & \textcolor{gray}{$2.4{\times}10^2$}   & Rigel; Blue Supergiant\\
        
        \hline
        &&&&F160W filter\\
        \hline
        
        G5V   & 1   & 1   &        1.00        & $5.750{\times}10^3$   & $5.6{\times}10^8$   & $8.8{\times}10^7$   & \textcolor{gray}{$1.4{\times}10^7$}   & Sun-like; Main-sequence\\
        B0V   & 15  & 6.5 & $2.04{\times}10^4$ & $3.000{\times}10^4$   & $3.3{\times}10^5$   & $5.3{\times}10^4$   & \textcolor{gray}{$8.4{\times}10^3$}   & Tau Scorpii; Main-sequence\\
        O3V   & 100 & 12  & $8.00{\times}10^5$ & $4.500{\times}10^4$   & $5.5{\times}10^4$   & $8.8{\times}10^3$   & \textcolor{gray}{$1.4{\times}10^3$}   & Main-sequence\\
        K5III & 1.2 & 45  & $4.40{\times}10^2$ & $4.000{\times}10^3$   & $2.0{\times}10^6$   & $3.2{\times}10^5$   & \textcolor{gray}{$5.1{\times}10^4$}   & Aldebaran; Red giant\\
        B8V   & 21  & 79  & $1.20{\times}10^5$ & $1.200{\times}10^4$   & $1.1{\times}10^4$   & $1.7{\times}10^3$   & \textcolor{gray}{$2.7{\times}10^2$}   & Rigel; Blue Supergiant\\
        \hline
    \end{tabular}
    \label{tab:magnitude}
    { \textbf{Note:} Spectral energy distributions and corresponding temperatures were taken from \citet{castelli2004new}. Nominal masses and temperatures were taken from \citet{2000ApJ...534..348H,2012ApJ...747..108M,2015A&A...580A..31H} and do not affect the calculation. Sizes and luminosities were taken from \citet{2000ApJ...534..348H, 2012ApJ...746..154P,2015A&A...580A..31H, 2015A&A...582A..49H,2012ApJ...747..108M}, to match the specific stars mentioned in the \emph{comments} column (in case mentioned). For Riegel-like (nominally Type B8Ia), blue supergiant we used for the calculation a B8V spectral energy distribution, but with the radius and luminosity specified herein, corresponding to Riegel. Similarly, we use a G5V spectrum for a sun-like star (nominally Type G2V), but incorporate solar values. Columns 6, 7, and 8 show the magnification required to observe the stars given an observational depth of 27 AB , 29, and 31 AB magnitudes, respectively.}
\end{table*}

\begin{table*}
    \centering
    \caption{Required magnification for different types of stars to be observe with JWST in various
    bands and for various depths.}
    \begin{tabular}{lllcccccccc}
        \hline
        Spectral type & Mass (${\rm M}_{\odot}$) & Size (${\rm R}_{\odot}$) & Luminosity (${\rm L}_{\odot}$) & Temperature (K) & & $\mu_{\rm c}$ & & comments \\
        & & & & & 27 AB & 29 AB & 31 AB & \\
        (1) & (2) & (3) & (4)&(5) & (6)& (7)&(8)&(9)\\
        
        \hline
        &&&&F090W filter\\
        \hline
        
        G5V   & 1   & 1   &        1.00        & $5.750{\times}10^3$   & $2.7{\times}10^9$   & $4.3{\times}10^8$   & $6.9{\times}10^7$   & Sun-like; Main-sequence\\
        B0V   & 15  & 6.5 & $2.04{\times}10^4$ & $3.000{\times}10^4$   & $1.6{\times}10^5$   & $2.5{\times}10^4$   & $3.9{\times}10^3$   & Tau Scorpii; Main-sequence\\
        O3V   & 100 & 12  & $8.00{\times}10^5$ & $4.500{\times}10^4$   & $2.3{\times}10^4$   & $3.6{\times}10^3$   & $5.7{\times}10^2$   & Main-sequence\\
        K5III & 1.2 & 45  & $4.40{\times}10^2$ & $4.000{\times}10^3$   & $1.1{\times}10^8$   & $1.8{\times}10^7$   & $2.9{\times}10^6$   & Aldebaran; Red giant\\
        B8V   & 21  & 79  & $1.20{\times}10^5$ & $1.200{\times}10^4$   & $1.2{\times}10^4$   & $1.9{\times}10^3$   & $3.0{\times}10^2$   & Rigel; Blue Supergiant\\
        
        \hline
        &&&&F115W filter\\
        \hline
        
        G5V   & 1   & 1   &        1.00        & $5.750{\times}10^3$   & $8.9{\times}10^8$   & $1.4{\times}10^8$   & $2.2{\times}10^7$   & Sun-like; Main-sequence\\
        B0V   & 15  & 6.5 & $2.04{\times}10^4$ & $3.000{\times}10^4$   & $2.1{\times}10^5$   & $3.3{\times}10^4$   & $5.2{\times}10^3$   & Tau Scorpii; Main-sequence\\
        O3V   & 100 & 12  & $8.00{\times}10^5$ & $4.500{\times}10^4$   & $3.3{\times}10^4$   & $5.2{\times}10^3$   & $8.2{\times}10^3$   & Main-sequence\\
        K5III & 1.2 & 45  & $4.40{\times}10^2$ & $4.000{\times}10^3$   & $7.3{\times}10^6$   & $1.2{\times}10^6$   & $1.8{\times}10^5$   & Aldebaran; Red giant\\
        B8V   & 21  & 79  & $1.20{\times}10^5$ & $1.200{\times}10^4$   & $7.8{\times}10^3$   & $1.2{\times}10^3$   & $2.0{\times}10^2$   & Rigel; Blue Supergiant\\
        
        \hline
        &&&&F150W filter\\
        \hline
        
        G5V   & 1   & 1   &        1.00        & $5.750{\times}10^3$   & $5.7{\times}10^8$   & $9.1{\times}10^7$   & $1.4{\times}10^7$   & Sun-like; Main-sequence\\
        B0V   & 15  & 6.5 & $2.04{\times}10^4$ & $3.000{\times}10^4$   & $3.2{\times}10^5$   & $5.0{\times}10^4$   & $8.0{\times}10^3$   & Tau Scorpii; Main-sequence\\
        O3V   & 100 & 12  & $8.00{\times}10^5$ & $4.500{\times}10^4$   & $5.3{\times}10^4$   & $8.3{\times}10^3$   & $1.3{\times}10^3$   & Main-sequence\\
        K5III & 1.2 & 45  & $4.40{\times}10^2$ & $4.000{\times}10^3$   & $2.2{\times}10^6$   & $3.4{\times}10^5$   & $5.4{\times}10^4$   & Aldebaran; Red giant\\
        B8V   & 21  & 79  & $1.20{\times}10^5$ & $1.200{\times}10^4$   & $1.0{\times}10^4$   & $1.6{\times}10^3$   & $2.6{\times}10^2$   & Rigel; Blue Supergiant\\
        
        \hline
        &&&&F200W filter\\
        \hline
        
        G5V   & 1   & 1   &        1.00        & $5.750{\times}10^3$   & $5.0{\times}10^8$   & $7.9{\times}10^7$   & $1.3{\times}10^7$   & Sun-like; Main-sequence\\
        B0V   & 15  & 6.5 & $2.04{\times}10^4$ & $3.000{\times}10^4$   & $5.2{\times}10^5$   & $8.3{\times}10^4$   & $1.3{\times}10^4$   & Tau Scorpii; Main-sequence\\
        O3V   & 100 & 12  & $8.00{\times}10^5$ & $4.500{\times}10^4$   & $8.8{\times}10^4$   & $1.4{\times}10^4$   & $2.2{\times}10^3$   & Main-sequence\\
        K5III & 1.2 & 45  & $4.40{\times}10^2$ & $4.000{\times}10^3$   & $1.2{\times}10^6$   & $1.9{\times}10^5$   & $3.0{\times}10^4$   & Aldebaran; Red giant\\
        B8V   & 21  & 79  & $1.20{\times}10^5$ & $1.200{\times}10^4$   & $1.5{\times}10^4$   & $2.3{\times}10^3$   & $3.7{\times}10^2$   & Rigel; Blue Supergiant\\
        
        \hline
        &&&&F277W filter\\
        \hline
        
        G5V   & 1   & 1   &        1.00        & $5.750{\times}10^3$   & $5.0{\times}10^8$   & $7.9{\times}10^7$   & $1.2{\times}10^7$   & Sun-like; Main-sequence\\
        B0V   & 15  & 6.5 & $2.04{\times}10^4$ & $3.000{\times}10^4$   & $9.3{\times}10^5$   & $1.5{\times}10^5$   & $2.3{\times}10^4$   & Tau Scorpii; Main-sequence\\
        O3V   & 100 & 12  & $8.00{\times}10^5$ & $4.500{\times}10^4$   & $1.6{\times}10^5$   & $2.5{\times}10^4$   & $4.0{\times}10^3$   & Main-sequence\\
        K5III & 1.2 & 45  & $4.40{\times}10^2$ & $4.000{\times}10^3$   & $7.7{\times}10^5$   & $1.2{\times}10^5$   & $1.9{\times}10^4$   & Aldebaran; Red giant\\
        B8V   & 21  & 79  & $1.20{\times}10^5$ & $1.200{\times}10^4$   & $2.1{\times}10^4$   & $3.4{\times}10^3$   & $5.3{\times}10^2$   & Rigel; Blue Supergiant\\
        
        \hline
        &&&&F356W filter\\
        \hline
        
        G5V   & 1   & 1   &        1.00        & $5.750{\times}10^3$   & $5.2{\times}10^8$   & $8.3{\times}10^7$   & $1.3{\times}10^7$   & Sun-like; Main-sequence\\
        B0V   & 15  & 6.5 & $2.04{\times}10^4$ & $3.000{\times}10^4$   & $1.5{\times}10^6$   & $2.4{\times}10^5$   & $3.8{\times}10^4$   & Tau Scorpii; Main-sequence\\
        O3V   & 100 & 12  & $8.00{\times}10^5$ & $4.500{\times}10^4$   & $2.6{\times}10^5$   & $4.2{\times}10^4$   & $6.6{\times}10^3$   & Main-sequence\\
        K5III & 1.2 & 45  & $4.40{\times}10^2$ & $4.000{\times}10^3$   & $5.7{\times}10^4$   & $9.1{\times}10^4$   & $1.4{\times}10^4$   & Aldebaran; Red giant\\
        B8V   & 21  & 79  & $1.20{\times}10^5$ & $1.200{\times}10^4$   & $3.2{\times}10^4$   & $5.2{\times}10^3$   & $8.1{\times}10^2$   & Rigel; Blue Supergiant\\
        
        \hline
        &&&&F410M filter\\
        \hline
        
        G5V   & 1   & 1   &        1.00        & $5.750{\times}10^3$   & $5.6{\times}10^8$   & $8.9{\times}10^7$   & $1.4{\times}10^7$   & Sun-like; Main-sequence\\
        B0V   & 15  & 6.5 & $2.04{\times}10^4$ & $3.000{\times}10^4$   & $2.0{\times}10^6$   & $3.2{\times}10^5$   & $5.0{\times}10^4$   & Tau Scorpii; Main-sequence\\
        O3V   & 100 & 12  & $8.00{\times}10^5$ & $4.500{\times}10^4$   & $3.5{\times}10^5$   & $5.5{\times}10^4$   & $8.7{\times}10^3$   & Main-sequence\\
        K5III & 1.2 & 45  & $4.40{\times}10^2$ & $4.000{\times}10^3$   & $4.9{\times}10^5$   & $7.7{\times}10^4$   & $1.2{\times}10^4$   & Aldebaran; Red giant\\
        B8V   & 21  & 79  & $1.20{\times}10^5$ & $1.200{\times}10^4$   & $4.1{\times}10^4$   & $6.4{\times}10^3$   & $1.0{\times}10^3$   & Rigel; Blue Supergiant\\
        
        \hline
        &&&&F444W filter\\
        \hline
        
        G5V   & 1   & 1   &        1.00        & $5.750{\times}10^3$   & $6.2{\times}10^8$   & $9.8{\times}10^7$   & $1.6{\times}10^7$   & Sun-like; Main-sequence\\
        B0V   & 15  & 6.5 & $2.04{\times}10^4$ & $3.000{\times}10^4$   & $2.3{\times}10^6$   & $3.6{\times}10^5$   & $5.7{\times}10^4$   & Tau Scorpii; Main-sequence\\
        O3V   & 100 & 12  & $8.00{\times}10^5$ & $4.500{\times}10^4$   & $3.9{\times}10^5$   & $6.3{\times}10^4$   & $9.9{\times}10^3$   & Main-sequence\\
        K5III & 1.2 & 45  & $4.40{\times}10^2$ & $4.000{\times}10^3$   & $5.1{\times}10^5$   & $8.0{\times}10^4$   & $1.3{\times}10^4$   & Aldebaran; Red giant\\
        B8V   & 21  & 79  & $1.20{\times}10^5$ & $1.200{\times}10^4$   & $4.5{\times}10^4$   & $7.2{\times}10^3$   & $1.1{\times}10^3$   & Rigel; Blue Supergiant\\
        
        \hline
    \end{tabular}
    \label{tab:magnitude_jwst}
    {\raggedright \textbf{Note:} Spectral energy distributions and corresponding temperatures were taken from \citet{castelli2004new}. Nominal masses and temperatures were taken from \citet{2000ApJ...534..348H,2012ApJ...747..108M,2015A&A...580A..31H} and do not affect the calculation. Sizes and luminosities were taken from \citet{2000ApJ...534..348H, 2012ApJ...746..154P,2015A&A...580A..31H, 2015A&A...582A..49H,2012ApJ...747..108M}, to match the specific stars mentioned in the \emph{comments} column (in case mentioned). For Riegel-like (nominally Type B8Ia), blue supergiant we used for the calculation a B8V spectral energy distribution, but with the radius and luminosity specified herein, corresponding to Riegel. Similarly, we use a G5V spectrum for a sun-like star (nominally Type G2V), but incorporate solar values. Columns 6, 7, and 8 show the magnification required to observe the stars given an observational depth of 27 AB , 29, and 31 AB magnitudes, respectively.}
\end{table*}

\bsp	
\label{lastpage}
\end{document}